\documentclass[aps,prd,preprintnumbers,showpacs,showkeys,nofootinbib,
superscriptaddress,fleqn,floatfix,tightenlines,10pt]{revtex4-1}
\usepackage{amsmath,amsfonts,amssymb,amscd,amsxtra,amsthm}
\usepackage{graphicx}   
\usepackage{epstopdf}
\usepackage{dcolumn}    
\usepackage{bm}         
\usepackage{slashed}
\usepackage{cancel}
\usepackage{float}
\usepackage{mathtools}
\usepackage{amsbsy}
\usepackage{amstext}

\usepackage[utf8]{inputenc}
\usepackage{booktabs}
\usepackage[normalem]{ulem} 
\usepackage[dvipsnames]{xcolor} 
\usepackage{tabularx}
\usepackage{enumitem}
\usepackage{array}
\usepackage{slashed}
\usepackage{tikz}
\usepackage{float}
\usepackage{multirow}
\renewcommand\sout{\bgroup \color{red} \ULdepth=-.5ex \ULset}

\begin{document}
\preprint{PKNU-NuHaTh-2019-03}
\title{Pomeron, nucleon-resonance, and $(0^+,0^-,1^+)$-meson \\ contributions
in $\phi$-meson
photoproduction}
\author{Sang-Ho Kim}
\email[E-mail: ]{sangho_kim@korea.ac.kr}
\affiliation{Center for Extreme Nuclear Matters (CENuM), Korea University,
Seoul 02841, Korea}
\affiliation{Department of Physics, Pukyong National University (PKNU),
Busan 48513, Korea}
\author{Seung-il Nam}
\email[E-mail: ]{sinam@pknu.ac.kr}
\affiliation{Department of Physics, Pukyong National University (PKNU),
Busan 48513, Korea}
\affiliation{Asia Pacific Center for Theoretical Physics (APCTP), Pohang
37673, Korea}
\date{\today}
\begin{abstract}
  We investigate the reaction mechanism of the $\phi$-meson photoproduction off the proton target, i.e., $\gamma p\to\phi p$, up to $\sqrt{s}=2.8$ GeV. For this purpose, we employ an effective Lagrangian approach in the tree-level Born approximation, and we employ various experimental and theoretical inputs. As a theoretical setup, the vectorlike Pomeron ($C=+1$) is taken into account as a parameterized two-gluon exchange contribution. We also consider $f_1(1285)$ axial-vector-meson, ($\pi,\eta$) pseudoscalar-meson, and ($a_0,f_0$) scalar-meson exchanges in the $t$ channel, in addition to the experimentally confirmed nucleon resonances, such as $N^*(2000,5/2^+)$ and $N^*(2300,1/2^+)$, for the direct $\phi$-meson radiations in the $s$ and $u$ channels. We provide numerical results for the total and differential cross sections as well as the spin-density matrices in the Gottfried-Jackson, Adair, and helicity frames. We observe that, together with the universally accepted pomeron contribution, the considered meson and nucleon-resonance contributions play significant roles in reproducing the experimental data for the forward and backward $\phi$-meson scattering-angle regions, respectively, indicating the nontrivial interferences between mesonic and baryonic contributions.
\end{abstract}
\pacs{13.60.Le, 13.40.-f, 14.20.Jn, 14.20.Gk}
\keywords{$\phi$-meson photoproduction, Pomeron, axial-vector meson, nucleon  resonance,
 effective Lagrangian approach.}
\maketitle

\section{Introduction}
$\phi$-meson photoproduction off the proton target, i.e., $\gamma p \to\phi p$, has attracted a lot of interest especially, since the bump structure was observed at very forward-angle regions near the threshold in the experiment of the LEPS Collaboration~\cite{Mibe:2005er}.
Because the $\phi$-meson is characterized by the hidden strangeness, hyperon
exchanges in the $u$ channel are forbidden and nucleon-pole contributions in the $s$ channel are also suppressed, due to
the Okubo-Zwieg-Iizuka (OZI) rule~\cite{Okubo:1963fa,Okubo:1963fa2,Okubo:1963fa3}.
Thus, the reaction mechanism of the $\phi$-meson photoproduction is distinguished from
those of the $\rho$- and $\omega$-meson photoproductions, resulting in its cross section being much smaller than others.
It is also well known that diffractive Pomeron ($\mathbb{P}$) exchange governs the monotonically increasing high-energy behavior of the cross section, but can not explain the bump structure near the threshold region $\sqrt{s}=(2.0-2.2)$ GeV. Refs.~\cite{Ozaki:2009mj,Ryu:2012tw}
interpret the bump structure as the coupled-channel effects between the
$\phi p$ and $K^+ \Lambda(1520)$ channels. Moreover, Refs.~\cite{Kiswandhi:2010ub,Kiswandhi:2011cq} suggest a
postulated spin-$3/2$ resonance with $M_{N^*} \approx 2.1$ GeV and $\Gamma_{N^*} \approx 500$ MeV to reproduce the bump. 

Meanwhile, in 2014, the CLAS Collaboration at Jefferson Laboratory~\cite{Seraydaryan:2013ija,
Dey:2014tfa} reported the first abundant data containing both the charged
and neutral modes of the $\phi \to K \bar K$ decay for the $\phi$-meson photoproduction.
These high-statistics differential cross sections and spin-density matrix
elements (SDMEs) data cover the energy range $\sqrt{s}=(2.0-2.8)$
GeV and the full angular range beyond the very forward angle.
They imply many interesting features:
i) The local structure studied previously persists only in the forward
angle regions and vanishes around the $\phi$-meson scattering angle
$\cos\theta \approx 0.8$ in the center-of-mass (c.m.) frame. Then, two
bumplike structures are shown again at backward angles
$\cos\theta=-(0.4-1.0)$ near $\sqrt{s} \approx 2.1$ and $2.3$ GeV,
although the magnitudes of the cross sections are far more suppressed than
those at forward angles.
ii) The comparison of the differential cross sections between the charged
($\phi \to K^+ K^-$) and neutral ($\phi \to K_S^0 K_L^0$)
modes~\cite{Dey:2014tfa} offers the best chance to extract the re-scattering
effect between the $\phi p$ and $K^+ \Lambda(1520)$ channels, because the
neutral mode excludes the $\Lambda (1520) \to p K^-$ final state
configuration.
The similarity between these two modes implies that the re-scattering effect and the interference are marginal. The bump structure at $\sqrt{s} \approx 2.2$ GeV is also clearly seen in the both modes.
Moreover, the LEPS Collaboration~\cite{Ryu:2016jmv} subsequently confirmed
that the $\sqrt{s} \approx 2.2$ GeV structure is regardless of the
$\phi$-$\Lambda(1520)$ interference effects in the $\gamma p \to K^+ K^- p$
reaction. Keeping these in mind, other reaction processes should come into play essentially~\cite{Dey:2014npa}. iii) The SDME $\rho_{00}^0$ provides us with the information of helicity conservation between the initial photon and the final $\phi$ meson, because it is proportional to the squares of the two helicity-flip amplitudes.
Nonzero values of $\rho_{00}^0$ are observed in the three different
reference frames, i.e., the Gottfried-Jackson, Adair, and helicity ones, resulting in violation of both $t$-channel helicity conservation and $s$-channel helicity conservation.
In this regard, it is of great importance to carry out a systematical
analysis of the $\phi$-meson photoproduction.

In the present work, we investigate the reaction mechanism of the $\phi$-meson
photoproduction with an effective Lagrangian approach in the tree-level Born
approximation.
To take into account the spatial distributions of hadrons involved,
well-established phenomenological form factors are considered as well.
We take into account the exchanges of the vectorlike Pomeron ($\mathbb{P}$),
the $f_1(1285)$ axial-vector (AV) meson, the ($\pi,\eta$) pseudoscalar (PS)
mesons, and the ($a_0,f_0$) scalar (S) mesons in the $t$-channel Feynman diagram.
THe general argument about the $\phi$-meson photoproduction is that the
conventional Pomeron exchange governs even at low energies and meson exchanges
are suppressed, because the OZI rule~\cite{Okubo:1963fa} puts constraints on
direct exchanges of quarks for the $\gamma\mathbb{P}\phi$ vertex.
By the analysis of the world CLAS data on the $\phi$-meson photoproduction, we
can readily test,
to what extent of energies, the Pomeron exchange has its effect on the cross
sections and SDMEs.
The relative contributions of the AV, PS, and S mesons also can be verified.
In addition, the direct $\phi$-meson radiations are considered in the $s$ and
$u$ channels, through the ground-state nucleon and its resonances.

Among the nucleon resonances given in the Particle Data Group (PDG)
~\cite{Tanabashi:2018oca}, we include $N^*(2000,5/2^+)$ and $N^*(2300,1/2^+)$, which are located near the $\phi N$ threshold.
We find that they are the most essential for describing the CLAS data, instead of
other scenarios such as a single hypothetical resonance~\cite{Ozaki:2009mj,
Kiswandhi:2010ub,Kiswandhi:2011cq}, meson-baryon box-shape loop
contributions~\cite{Ryu:2012tw}, and the interference of the $\phi$ meson with
$\Lambda(1520)$ via $\gamma p\to K^+K^-p$~\cite{Ryu:2016jmv}.
We face some difficulty owing to the lack of information about the $N^* \to
\phi N$ decays of the
PDG resonances, while their photo-excitations
$\gamma N \to N^*$ are relatively well known.
We extract the branching ratios for $N^*\to \phi N$ by fitting to the
CLAS data and compare them with those for the open-strangeness $N^* \to K^*
\Lambda$ decays recently reported by the PDG~\cite{Tanabashi:2018oca}.

This paper is organized as follows: In Sec.~\ref{SecII}, we present
detailed explanations for the theoretical framework.
Section~\ref{SecIII} is devoted to the numerical results of cross sections
and SDMEs and the relevant discussions.
The summary is given in Sec.~\ref{SecIV}. The details of the invariant amplitudes,
SDMEs, and three reference frames are given in the appendixes.

\section{Theoretical Framework}
\label{SecII}
\begin{figure}[b]
\includegraphics[width=6cm]{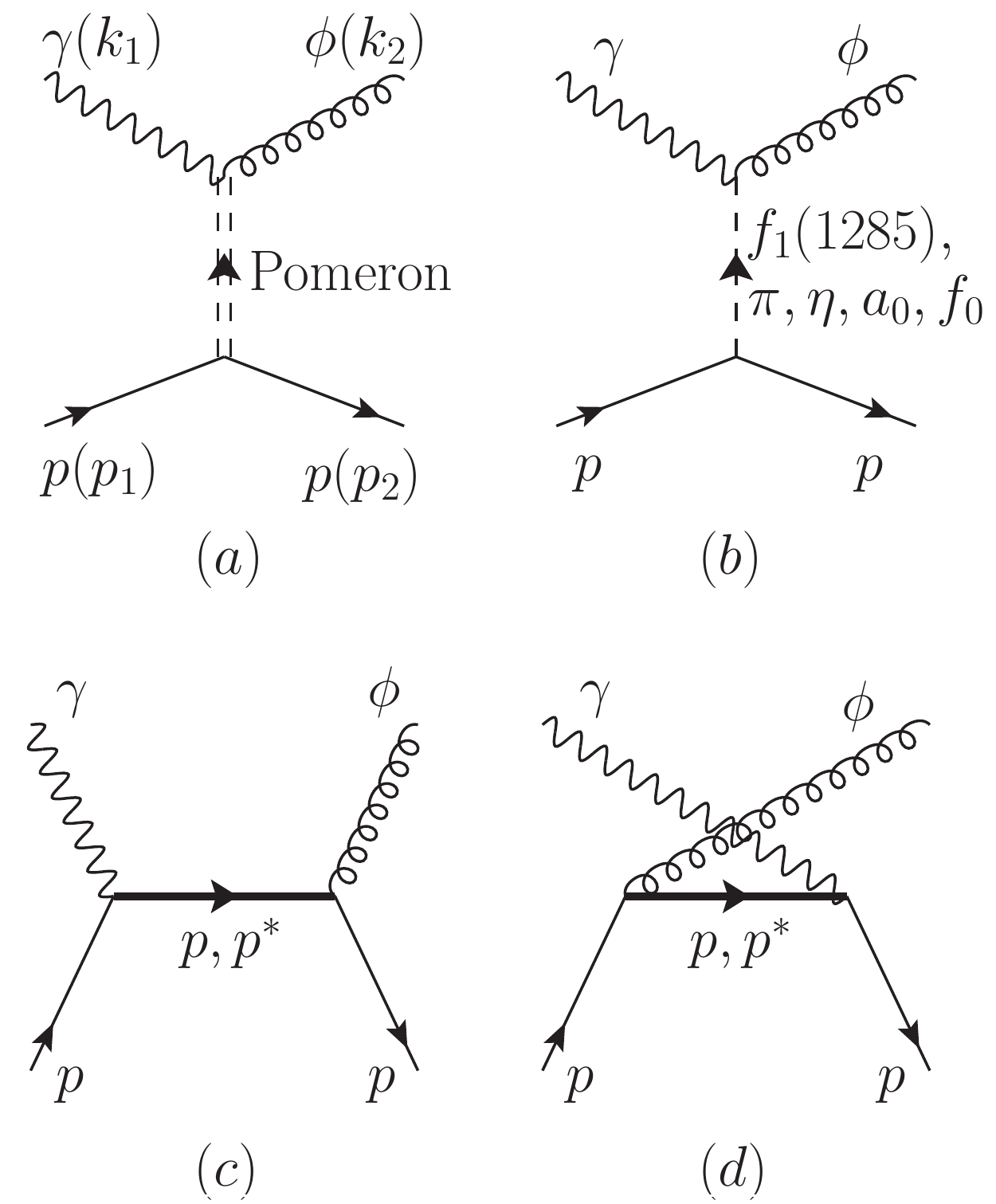}
\caption{Relevant Feynman diagrams for $\gamma p\to \phi p$, which include
Pomeron ($a$), pseudoscalar ($\pi,\eta$)-meson , scalar ($a_0,f_0$)-meson, and
axial-vector [$f_1(1285)$]-meson exchanges in the $t$ channel ($b$), and direct
$\phi$-meson
radiations via the proton and its resonances in the $s$ and $u$ channels ($c$ and
$d$).
We define the four momenta for the involved particles as well.}
\label{FIG1_FD}
\end{figure}

In this section, we provide the theoretical framework to study the $\phi$-meson
photoproduction off the proton target $\gamma (k_1) + p (p_1) \to \phi (k_2) + p
(p_2)$.
We employ an effective Lagrangian approach and the tree-level Feynman diagrams
under consideration are depicted in Fig.~\ref{FIG1_FD}. The Reggeized two-gluon
exchange, i.e., the Pomeron ($\mathbb{P}$), is taken into account in the $t$
channel to describe the slowly rising total cross section with respect to the
beam energy [Fig.~\ref{FIG1_FD}(a)].
Exchanges of the $f_1(1285)$ AV, ($\pi,\eta$) PS, and ($a_0,f_0$) S mesons are
considered in the $t$ channel [Fig.~\ref{FIG1_FD}(b)].
The direct $\phi$-meson radiations in the $s$ and $u$ channels via the proton
and its resonances are also available
[Figs.~\ref{FIG1_FD}(c) and~\ref{FIG1_FD}(d)].
In what follows, the explicit forms of the effective Lagrangians are
explained for describing various hadron interactions for each Feynman diagram.

\subsection{Vectorlike Pomeron  exchange}
\begin{figure}[htp]
\includegraphics[width=4.5cm]{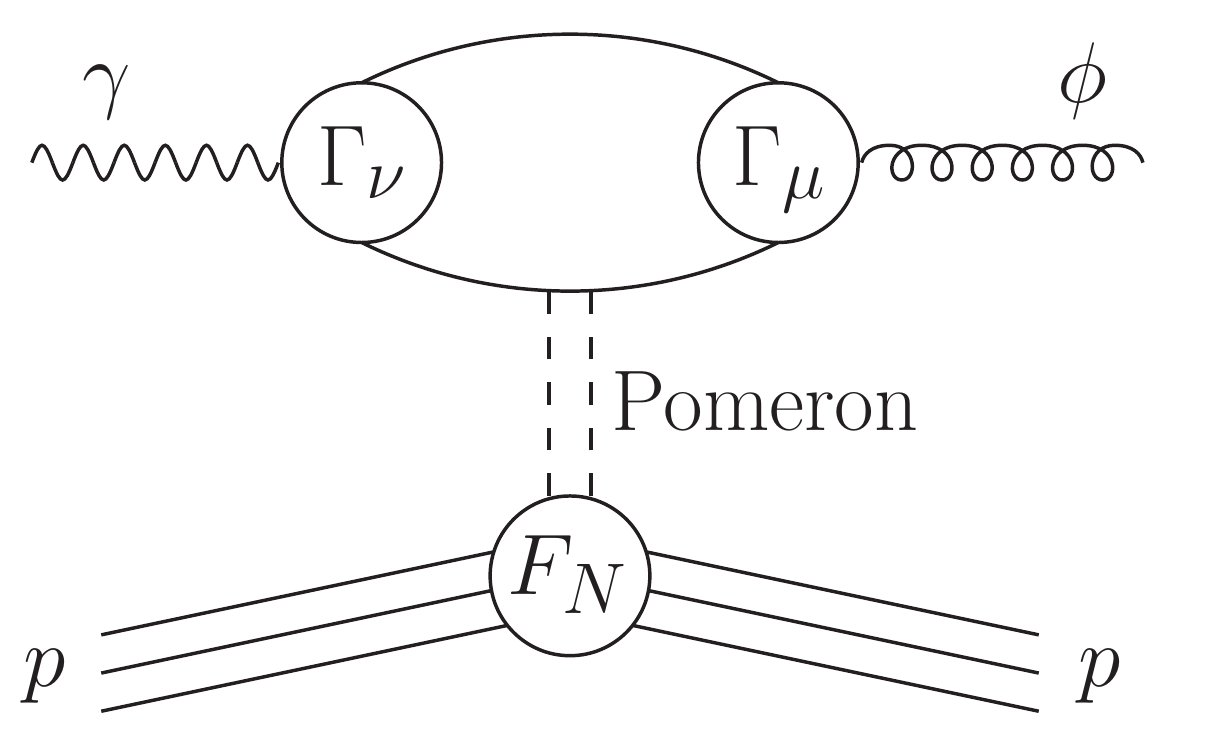}
\caption{Quark diagram for Pomeron exchange in the DL model.}
\label{FIG2_QD}
\end{figure}

The effective Lagrangians for Pomeron exchange for the
photon-Pomeron-$\phi$-meson vertex can be written
as~\cite{Titov:1999eu,Titov:2003bk}
\begin{align}
\label{eq:Lag:Pom1}
\mathcal{L}_{\gamma\mathbb{P}\phi} = ig_{\gamma\mathbb{P}\phi} F_\phi(t)
\left[\left( \phi_\nu \partial^\mu \mathbb{P}^\nu - \mathbb{P}_\nu \partial^\mu
\phi^\nu \right) A_\mu -
\left( A_\nu\partial^\mu \mathbb{P}^\nu - \mathbb{P}_\nu \partial^\mu A^\nu
\right) \phi_\mu -
\left( \phi_\nu \partial^\mu A^\nu - A_\nu \partial^\mu \phi^\nu \right)
\mathbb{P}_\mu \right] ,
\end{align}
where, $\mathbb{P}^\mu$, $A^\mu$, and $\phi^\mu$ indicate the
$C=+1$ vector like Pomeron, photon, and $\phi$-meson fields, respectively.
Considering its vector like nature, the interaction of the Pomeron with the
nucleon is casted into
\begin{align}
\label{eq:Lag:Pom2}
\mathcal{L}_{\mathbb{P}NN} =
g_{\mathbb{P}NN} F_N(t) \bar N \gamma_\mu N \mathbb{P}^\mu.
\end{align}
Here, $N$ stands for the nucleon field. We define the strength $C_\mathbb{P}$
for convenience by $C_\mathbb{P}= g_{\gamma\mathbb{P}\phi}\,g_{\mathbb{P}NN}$.
The form of the invariant amplitude derived from Eqs.~(\ref{eq:Lag:Pom1})
and~(\ref{eq:Lag:Pom2}) is similar to that used in the Donnachie-Landshoff (DL)
model~\cite{Titov:1999eu,Titov:2003bk,Donnachie:1987abc} given in Appendix A.
In this work, we employ the DL model for the Pomeron exchange process.
The complicated two-gluon exchange calculation is simplified by the DL
model which suggests that the Pomeron couples to the nucleon like a $C=+1$
isoscalar photon and its couplings is described in terms of a nucleon
isoscalar electromagnetic (EM) form factor $F_N(t)$.
Figure~\ref{FIG2_QD} draws the quark diagram for this Reggeized Pomeron exchange.
In Refs.~\cite{Pichowsky:1996jx,Pichowsky:1996tn}, the quark-loop
integration in Fig.~\ref{FIG2_QD} is performed in connection with the DL
model and can be approximated to the factorized form given by Donnachie and
Landshoff~\cite{Donnachie:1987abc} for the low energy region $\sqrt{s} < 5$ GeV,
justifying the application of the DL model to the present work.

The form factor for the $\gamma\mathbb{P}\phi$ vertex [$F_\phi(t)$]
~\cite{Donnachie:1988nj,Laget:1994ba} and the $F_N(t)$
~\cite{Jaroszkiewicz:1974ep,Donnachie:1983hf} read
\begin{align}
\label{eq:FF:Pom}
F_\phi(t)=\frac{2 \mu_0^2}{(1-t/\Lambda_\phi^2)(2\mu_0^2 + \Lambda_\phi^2 -t)},
\,\,\,\,\,
F_N(t) = \frac{4M^2_N- a_N^2 t}{(4M^2_N-t)(1-t/t_0)^2},
\end{align}
respectively, where the scale parameters are given by $\mu_0^2 = 1.1\,
\mathrm{GeV}^2$, $a_N^2=2.8$, and $t_0= 0.71\,\mathrm{GeV}^2$.
The mass scale $\Lambda_\phi$ is proportional to the quark mass of the loop
diagram in Fig.~\ref{FIG2_QD} and is chosen to be $\Lambda_\phi^2 = 4 M_\phi^2$
which is rather larger than that given in Refs.~\cite{Titov:1999eu,Titov:2003bk,
Laget:1994ba} where $\Lambda_\phi^2 = M_\phi^2$ is used.
This modification makes the differential cross sections milder under the
variations of the scattering angle $\cos\theta$.
The Pomeron trajectory reads
\begin{align}
\label{eq:Pom:Tra}
\alpha_\mathbb{P}(t) = 1 + \epsilon_{\mathbb P} + \alpha'_{\mathbb P} t ,
\end{align}
where the slope and intercept of the trajectory are determined to be
$\alpha'_{\mathbb P} = 0.25\,\mathrm{GeV}^{-2}$ and $\epsilon_{\mathbb P}= 0.08$
which is favored among $ 0.08-0.12$~\cite{Donnachie:1983hf}.

\subsection{$f_1(1285)$ axial-vector exchange}
The $f_1(1285)$ meson exchange is suggested in proton-proton scattering and
vector meson photoproduction due to its special relation to the axial anomaly
through the matrix elements of the flavor singlet axial vector current.
Its Regge trajectory is expected to contribute to the large-energy and
large-momentum transfer regions with an intercept of $\alpha (0) \approx 1$ and
a slope of $\alpha' \approx 0$ as the odd-signature partner of the even
signature Pomeron ~\cite{Kochelev:1999zf}.

The effective Lagrangian for the $AVV$ vertex is obtained by using the hidden
gauge approach~\cite{Kaiser:1990yf}:
\begin{align}
\label{eq:Lag:GPhif1}
\mathcal{L}_{\gamma \phi f_1} = g_{\gamma \phi f_1}
\epsilon^{\mu\nu\alpha\beta} \partial_\mu A_\nu
\partial^\lambda \partial_\lambda \phi_\alpha f_{1\beta} ,
\end{align}
where $f_1$ indicates the $f_1(1285)$ field with its quantum number $I^G (J^{PC})
= 0^+ (1^{++})$.
The coupling constant $g_{\gamma \phi f_1}$ can be calculated from the relation
\begin{align}
\label{eq:Wid:GPhif1}
\Gamma_{f_1 \to \phi \gamma} = \frac{k_\gamma^3}{12\pi} \frac{M_\phi^2}{M_{f_1}^2}
(M_{f_1}^2+ M_\phi^2) g_{\gamma \phi f_1}^2 ,
\end{align}
derived from Eq.~(\ref{eq:Lag:GPhif1}), where $k_\gamma =
(M_{f_1}^2- M_\phi^2)/(2M_{f_1})$, and the experimental data on the $f_1$-meson
branching ratio (Br) ${\mathrm{Br}}_{f_1 \to \phi \gamma} = 7.5 \times 10^{-4}$ with
$\Gamma_{f_1}$ = 22.7 MeV~\cite{Tanabashi:2018oca}:
\begin{align}
\label{eq:CC:GPhif1}
g_{\gamma \phi f_1} = 0.17 \, \mathrm{GeV^{-2}} .
\end{align}

The axial-vector meson and nucleon interaction Lagrangian reads
\begin{align}
\label{eq:Lag:f1NN}
\mathcal{L}_{f_1 NN} = - g_{f_1 NN} \bar N
\left[ \gamma_\mu - i \frac{\kappa_{f_1 NN}}{2M_N} \gamma_\nu \gamma_\mu
\partial^\nu \right] f_1^\mu \gamma_5 N,
\end{align}
where the coupling constant $g_{f_1NN}$ is not well known, and we use the maximum value
\begin{align}
\label{eq:CC:f1NN}
g_{f_1NN} = 3.0,
\end{align}
with $g_{f_1NN} =2.5 \pm 0.5$, discussed in Ref.~\cite{Birkel:1995ct}.
Although the tensor term can contribute to the $\phi$-meson photoproduction, we
set the value of $\kappa_{f_1 NN}$ to be zero in this calculation for brevity.

We reggeize the Feynman amplitude derived from Eqs.~(\ref{eq:Lag:GPhif1}) and
(\ref{eq:Lag:f1NN}) (see Appendix~A) by replacing the Feynman propagator by the
Regge propagator, which effectively interpolates between small- and large-momentum
transfer regions, such that the mesons of higher spin $J=3,5,\cdots$ in
the same trajectory can contribute to the high energy region~\cite{Donachie2002}
\begin{eqnarray}
\label{eq:REGGEPRO}
P_{f_1}^{\mathrm{Feyn}}(t) = \frac{1}{t-M_{f_1}^2} \to
P_{f_1}^{\mathrm{Regge}}(t) = \left(\frac{s}{s_{f_1}} \right)^{\alpha_{f_1}(t)-1}
\frac{\pi\alpha'_{f_1}}{\sin[\pi\alpha_{f_1}(t)]}
\frac{1}{\Gamma [\alpha_{f_1}(t)]} D_{f_1}(t) ,
\end{eqnarray}
where the energy-scale factor is fixed to be $s_{f_1} = 1\,\mathrm{GeV^2}$ and the
odd signature factor is given by
\begin{eqnarray}
\label{eq:SigFac}
D_{f_1}(t) = \frac{\mathrm{exp}[-i\pi\alpha_{f_1}(t)]-1}{2} .
\end{eqnarray}
The slope of the $f_1(1285)$ trajectory is chosen to be $\alpha_{f_1}' \approx
0.028\, \mathrm{GeV^{-2}}$, which is distinguished from that of the Pomeron
because of its different characteristic scale relative to the Pomeron
~\cite{Kochelev:1999zf}.
The intercept of the $f_1(1285)$ trajectory is determined to be $\alpha_{f_1}
(0) =0.99 \pm 0.04$~\cite{Kochelev:1999zf}, which is rather larger than those of
other vector and axial-vector meson trajectories~\cite{Brisudova:1999ut}.
\subsection{Pseudoscalar- and scalar-meson exchanges}
The EM interaction Lagrangians for the PS- and S-meson exchanges,
respectively, read
\begin{align}
\label{eq:Lag:GMPhi}
\mathcal{L}_{\gamma\Phi\phi} =& \frac{eg_{\gamma\Phi\phi}}{M_\phi}
\epsilon^{\mu\nu\alpha\beta} \partial_\mu A_\nu \partial_\alpha \phi_\beta \Phi ,\cr
\mathcal{L}_{\gamma S \phi} =& \frac{eg_{\gamma S \phi}}{M_\phi}
F^{\mu\nu} \phi_{\mu\nu} S ,
\end{align}
where $\Phi=\pi^0(135,0^-)$ and $\eta(548,0^-)$, and $S=a_0(980,0^+)$ and
$f_0(980,0^+)$.
$e$ stands for the unit electric charge.
The EM and $\phi$-meson field strengths are denoted by $F^{\mu\nu} = \partial^\mu
A^\nu - \partial^\nu A^\mu$ and $\phi^{\mu\nu} = \partial^\mu \phi^\nu - \partial^\nu
\phi^\mu$, respectively.
The relevant coupling constants are calculated from the widths of the $\phi \to
\Phi \gamma$ and $\phi \to S \gamma$ radiative decays as follows:
\begin{align}
\label{eq:Wid:GMPhi}
\Gamma_{\phi \to \Phi \gamma} =
\frac{\alpha}{3} \frac{q_\gamma^3}{M_\phi^2} g_{\gamma\Phi\phi}^2,\,\,\, 
\Gamma_{\phi \to S \gamma} =
\frac{4\alpha}{3} \frac{q_\gamma^3}{M_\phi^2} g_{\gamma S \phi}^2 ,
\end{align}
where $\alpha = e^2/(4\pi)$ and $q_\gamma = (M_\phi^2- M_{\Phi,S}^2)/(2M_\phi)$.
The $\phi$-meson branching ratios are experimentally known to be 
${\mathrm{Br}}_{\phi \to \pi \gamma} = 1.30 \times 10^{-3}$,
${\mathrm{Br}}_{\phi \to \eta \gamma} = 1.303 \times 10^{-2}$,
${\mathrm{Br}}_{\phi \to a_0 \gamma} = 7.6 \times 10^{-5}$, and
${\mathrm{Br}}_{\phi \to f_0 \gamma} = 3.22 \times 10^{-4}$~\cite{Tanabashi:2018oca},
from which we obtain
\begin{align}
\label{eq:CC:GMPhi}
g_{\gamma\pi\phi}=-0.14, \hspace{1em} g_{\gamma\eta\phi}=-0.71, \hspace{1em}
g_{\gamma a_0 \phi}= -0.77, \hspace{1em} g_{\gamma f_0 \phi}= -2.44 ,
\end{align}
with $\Gamma_\phi = 4.249$ MeV.

The strong interaction Lagrangians for the PS- and S-meson exchanges
are written as
\begin{align}
\label{eq:Lag:MNN}
\mathcal{L}_{\Phi NN} =& -ig_{\Phi NN} \bar N \Phi \gamma_5 N ,                \cr
\mathcal{L}_{S NN} =& -g_{S NN} \bar N SN ,
\end{align}
respectively.
We use the pseudoscalar meson-baryon coupling scheme rather than the
pseudovector one for the former one in Eq.~(\ref{eq:Lag:MNN}).
They are equivalent to each other because the relevant two nucleons are on
mass-shell.
The following strong coupling constants are obtained by using the Nijmegen
potentials~\cite{Stoks:1999bz,Stoks:1999bz2}:
\begin{align}
\label{eq:CC:MNN}
g_{\pi NN}= 13.0, \hspace{1em} g_{\eta NN}= 6.34,   \hspace{1em}
g_{a_0 NN}= 4.95, \hspace{1em} g_{f_0 NN}= -0.51,
\end{align}
which are also close to the values from the SU(3) flavor symmetry or the unitary
symmetry except for $g_{a_0 NN}$~\cite{Titov:1999eu}.

We use the following parametrization of the form factors for the PS- and
S-meson exchanges:
\begin{align}
\label{eq:FF:M}
F_{\Phi,S} (t) = \frac{\Lambda_{\Phi,S}^2 - M_{\Phi,S}^2}{\Lambda_{\Phi,S}^2 - t} ,
\end{align}
where $\Lambda_{\Phi,S}$ denotes the cutoff masses, which will be determined to
reproduce experimental data.

\subsection{Direct $\phi$-meson radiation term}
The effective Lagrangians for the direct $\phi$-meson radiation contributions
are defined by
\begin{align}
\label{eq:LAG:N}
\mathcal{L}_{\gamma NN} =& - e \bar N
\left[ \gamma_\mu - \frac{\kappa_N}{2M_N} \sigma_{\mu\nu} \partial^\nu
\right] N A^\mu ,                                                      \cr
\mathcal{L}_{\phi NN} =& - g_{\phi NN} \bar N
\left[ \gamma_\mu - \frac{\kappa_{\phi NN}}{2M_N} \sigma_{\mu\nu} \partial^\nu
\right] N \phi^\mu,
\end{align}
where the anomalous magnetic moment of the proton is given by
$\kappa_p = 1.79$.
The vector and tensor coupling constants for the $\phi$-meson to the
nucleon are chosen to be $g_{\phi NN}=-0.24$ and
$\kappa_{\phi NN}=0.2$~\cite{Meissner:1997qt}.
Note that the corresponding individual invariant amplitudes, given by
$\mathcal{M}_{\phi\,\mathrm{rad},s}$ and $\mathcal{M}_{\phi\,\mathrm{rad},u}$ in
Eq.~(\ref{eq:Amp:N}) in the Appendix~A, violate the Ward-Takahashi identity (WTI).
When we sum the electric terms of the two invariant amplitudes, the WTI is
restored as a pair.
Thus one needs a specific prescription for the usage of the phenomenological
form factors.
Detailed explanations for this can be found in Refs.~\cite{Ohta:1989ji,
Haberzettl:1997jg,Haberzettl:1998eq,Davidson:2001rk,Haberzettl:2015exa}.
We define the form factor as follows:
\begin{align}
\label{eq:FF:N}
F_N (x) = \frac{\Lambda^4_N}{\Lambda^4_N+(x-M^2_N)^2}, \,\,\,x=(s,u),
\end{align}
and we take the common form factor which conserves the on-shell condition for
the form factors~\cite{Davidson:2001rk} as
\begin{align}
\label{eq:FF:C}
F_c (s,u) = 1 - [1-F_N(s)][1-F_N(u)],
\end{align}
for the electric terms. Because the magnetic terms satisfy the WTI by
themselves, we just use the form of Eq.~(\ref{eq:FF:N}).

\subsection{Nucleon resonances}
There are 11 nucleon resonances beyond the $\phi N$ threshold
$\sqrt{s_{\phi N}} = 1.96$ GeV in the PDG data~\cite{Tanabashi:2018oca}:
$N^*(2000,5/2^+)$, $N^*(2060,5/2^-)$, $N^*(2100,1/2^+)$, $N^*(2120,3/2^-)$,
$N^*(2190,7/2^-)$, $N^*(2220,9/2^+)$, $N^*(2250,9/2^-)$, $N^*(2300,1/2^+)$,
$N^*(2570,5/2^-)$, $N^*(2600,11/2^-)$, and $N^*(2700,13/2^+)$
with their two-, three-, or four-star confirmations.
Whereas the helicity amplitudes of $N^* \to N \gamma$ transitions are well
known for the nucleon resonances less than 2.3 GeV, the information of $N^* \to
\phi N$ strong decay is very limited for all PDG nucleon resonances.
None of the $N^* \to \phi N$ decay is observed firmly experimentally
~\cite{Tanabashi:2018oca}.
And the CLAS data for the differential cross sections exhibit two bumplike
structures at the pole positions $\sqrt{s} \approx 2.1$ and $2.3$ GeV in the
backward-scattering regions~\cite{Dey:2014tfa}.
Considering these observations, we avoid performing a $\chi^2$ fit in this
calculation and try to reproduce the available experimental data with only a
few resonances.
As is explained in the next section in detail, we find that selecting only
$N^*(2000,5/2^+)$ and $N^*(2300,1/2^+)$ is good enough for this exploratory work.
The nucleon resonances beyond 2.3 GeV and with higher spins, i.e., $J \geq 9/2$,
are automatically excluded.

The effective Lagrangians for the EM transitions of the nucleon resonances
read~\cite{Oh:2011,Kim:2012pz}
\begin{align}
\label{eq:Lag:GNNs}
\mathcal{L}^{1/2^\pm}_{\gamma  N N^*} &=
\frac{eh_1}{2M_N} \bar N \Gamma^{(\mp)}
\sigma_{\mu\nu} \partial^\nu A^\mu N^* + \mathrm{H.c.} ,
\cr
\mathcal{L}^{5/2^\pm}_{\gamma N N^*}&=
e \left[ \frac{h_1}{(2M_N)^2} \bar N \Gamma_\nu^{(\mp)}
 - \frac{ih_2}{(2M_N)^3} \partial_\nu \bar N
 \Gamma^{(\mp)} \right] \partial^\alpha F^{\mu\nu} N^*_{\mu\alpha} + \mathrm{H.c.},
\end{align}
for their spin and parity $J^P$, and we define the following notation for
brevity:
\begin{align}
\label{eq:GammaPM}
\Gamma^{(\pm)} = \left(
\begin{array}{c}
\gamma_5 \\ I_{4\times4}
\end{array} \right),  \,\,\,\,
\Gamma_\mu^{(\pm)} = \left(
\begin{array}{c}
\gamma_\mu \gamma_5 \\ \gamma_\mu
\end{array} \right).
\end{align}
$N^*$ and $N^*_{\mu\alpha}$ stand for the Rarita-Schwinger spin-1/2 and -5/2
nucleon resonance fields, respectively~\cite{Rarita:1941mf}.
The former one in Eq.~(\ref{eq:Lag:GNNs}) is constructed from the $\gamma NN$
interaction in Eq.~(\ref{eq:LAG:N}), but the electric term is removed to conserve
the WTI.
The Breit-Wigner helicities $A_i$ for $N^*(2000,5/2^+)$ are given in the
PDG data~\cite{Tanabashi:2018oca} by $A_{1/2}^{N^*(2000) \to p \gamma}$ = $0.031$ and
$A_{3/2}^{N^*(2000) \to p \gamma}$ = $-0.043$ [$\mathrm{GeV}^{-\frac{1}{2}}$] and we
extract the EM transition coupling constants $h_i$ in Eq.~(\ref{eq:Lag:GNNs})
~\cite{Oh:2011,Shklyar:2004ba,Oh:2007jd} from these values: $h_1^{N^*(2000)}$ =
$-4.24$ and $h_2^{N^*(2000)}$ = $4.00$.

The effective Lagrangians for the strong interactions can be expressed as
\begin{align}
\label{eq:Lag:PhiNNs}
\mathcal{L}^{1/2^\pm}_{\phi N N^*} &=
\pm \frac{1}{2M_N} \bar N
\left[ \frac{g_1 M_\phi^2}{M_{N^*} \mp M_N} \Gamma_\mu^{(\mp)} \pm g_2
\Gamma_\mu^{(\mp)} \sigma_{\mu\nu} \partial^\nu \right] \phi^\mu N^* +
\mathrm{H.c.},                                                             \cr
\mathcal{L}^{5/2^\pm}_{\phi N N^*} &=
\left[ \frac{g_1}{(2M_N)^2} \bar N \Gamma_\nu^{(\mp)}
 - \frac{ig_2}{(2M_N)^3} \partial_\nu \bar N \Gamma^{(\mp)}
 + \frac{ig_3}{(2M_N)^3} \bar N \Gamma^{(\mp)} \partial_\nu \right]
\partial^\alpha \phi^{\mu\nu} N_{\mu\alpha}^* + \mathrm{H.c.} ,
\end{align}
Here, only the first term $g_1$ is considered to avoid additional ambiguities
and the values of $g_2$ and $g_3$ are set to be zero.
We determine the value of $g \equiv g_1$ very carefully to reproduce the CLAS
data, resulting in $g_{N^*(2000)} = 4.0$, and extract the corresponding branching
ratio: $\mathrm{Br}_{N^*(2000) \to \phi N} = 1.5 \times 10^{-3}$.

It is worth comparing the computed branching ratio $\mathrm{Br}_{N^* \to \phi N}$
with those for the open strangeness $N^* \to K^* \Lambda$ decays.
The recent Bonn-Gatchina partial wave analysis of the $\gamma p \to K^{*+}
\Lambda$ reaction derived them, where $N^*(1895,1/2^-)$, $N^*(2000,5/2^+)$, and
$N^*(2100,1/2^+)$ turn out to be dominant and other several ones also give $N^*
\to K^* \Lambda$ branching ratios with small but finite values
~\cite{Anisovich:2017rpe}.
We obtain
\begin{align}
\label{eq:Br:Compare}
\frac{\mathrm{Br}(N^*(2000) \to \phi N)}
{\mathrm{Br}(N^*(2000) \to K^* \Lambda)} \simeq 0.07 .
\end{align}
For the case of $N^*(2300,1/2^+)$, because its photo-coupling is unknown, a
product of the EM and strong decay channels is taken into account to reproduce
the data as follows:
\begin{align}
\label{eq:BrA}
\sqrt{\mathrm{Br}_{N^*(2300) \to \phi N}} \times A_{1/2}^{N^*(2300) \to p \gamma} =
3.8 \times 10^{-3}\, \mathrm{GeV}^{-\frac{1}{2}} .
\end{align}
The full decay widths of the two nucleon resonances are chosen to be
$\Gamma_{N^*(2000)}$ = 200 MeV and $\Gamma_{N^*(2300)}$ = 300 MeV.

Meanwhile, Refs.~\cite{Zhao:1999af,Zhao:2001ue} used an SU(3) quark model to
probe the nucleon resonances that strongly couple to the $\phi N$ channel.
Reference~\cite{Lebed:2015dca} suggested the effect of the hidden-strangeness
pentaquark state $P_s^+ = s\bar s uud$ as its charmed partner, i.e., two
exotic charmoniumlike states $P^+_c(4312)$ and $P^+_c(4450)$, observed at
the LHCb Collaboration~\cite{Aaij:2015tga,Aaij:2019vzc}, are studied in the
same $s$-channel diagram for the $\gamma p \to J/\psi p$
photoproduction~\cite{Kubarovsky:2015aaa,Wang:2015jsa,Karliner:2015voa,
Blin:2016dlf}.

The Gaussian form factor is used because it is advantageous to suppress
unreasonably-increasing cross sections with respect to $\sqrt{s}$ for the $N^*$
contributions~\cite{Kim:2017nxg,Kim:2018qfu,Suh:2018yiu,Corthals:2005ce,
DeCruz:2012bv}:
\begin{align}
\label{eq:FF:Gau}
F_{N^*}(x) = \mathrm{exp}
\left[ - \frac{(x-M_{N^*}^2)^2}{\Lambda_{N^*}^4} \right],\,\,\,x=(s,u).
\end{align}
We observe that the contribution of the $u$-channel diagram of
Fig.~\ref{FIG1_FD}($d$) is almost negligible and all the $N^*$ contributions
come from the $s$-channel diagram of Fig.~\ref{FIG1_FD}($c$). 

\section{Numerical results and Discussions}
\label{SecIII}
\begin{figure}[htp]
\centering
\includegraphics[width=7.0cm]{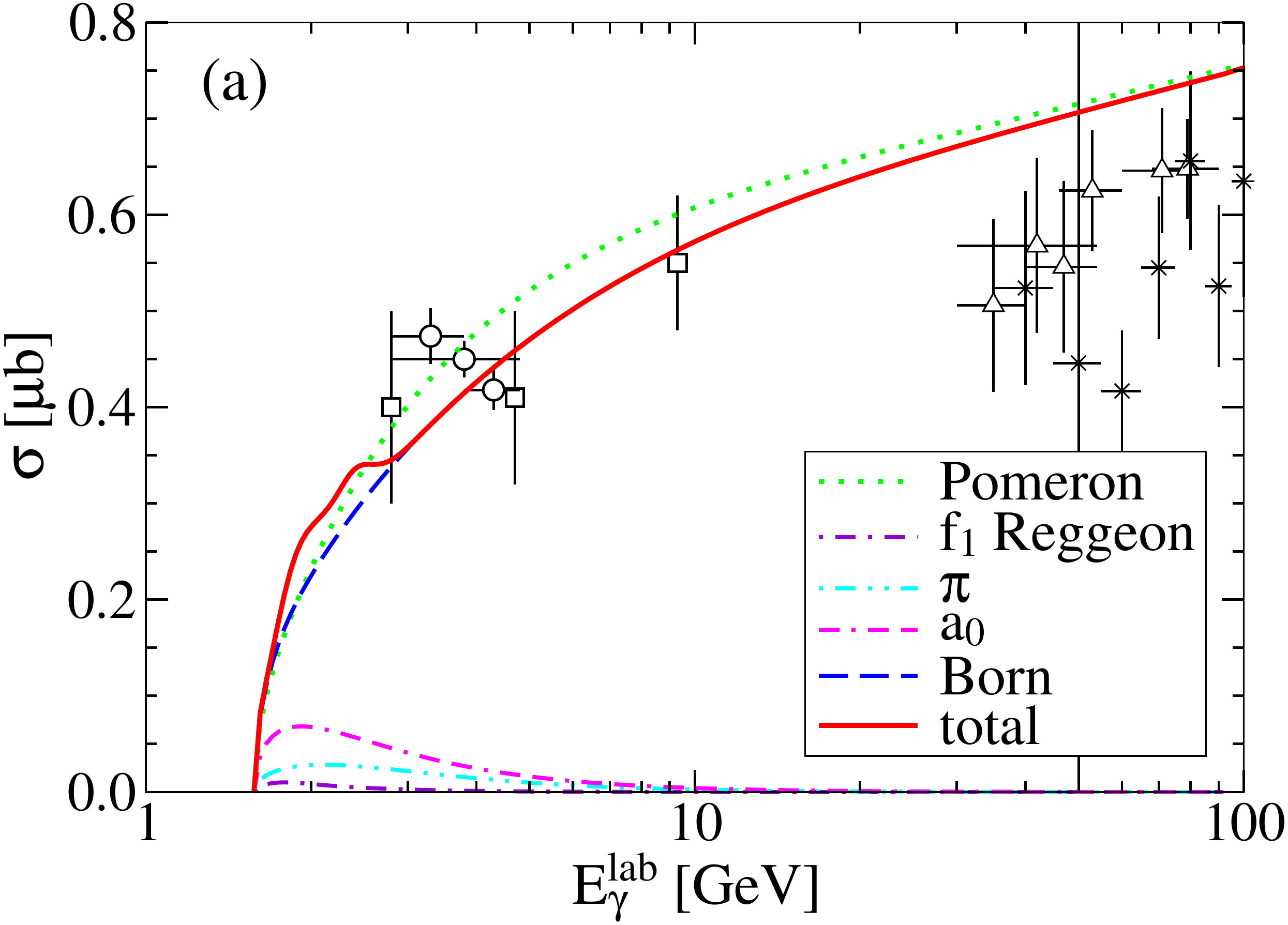} \hspace{1em}
\includegraphics[width=7.0cm]{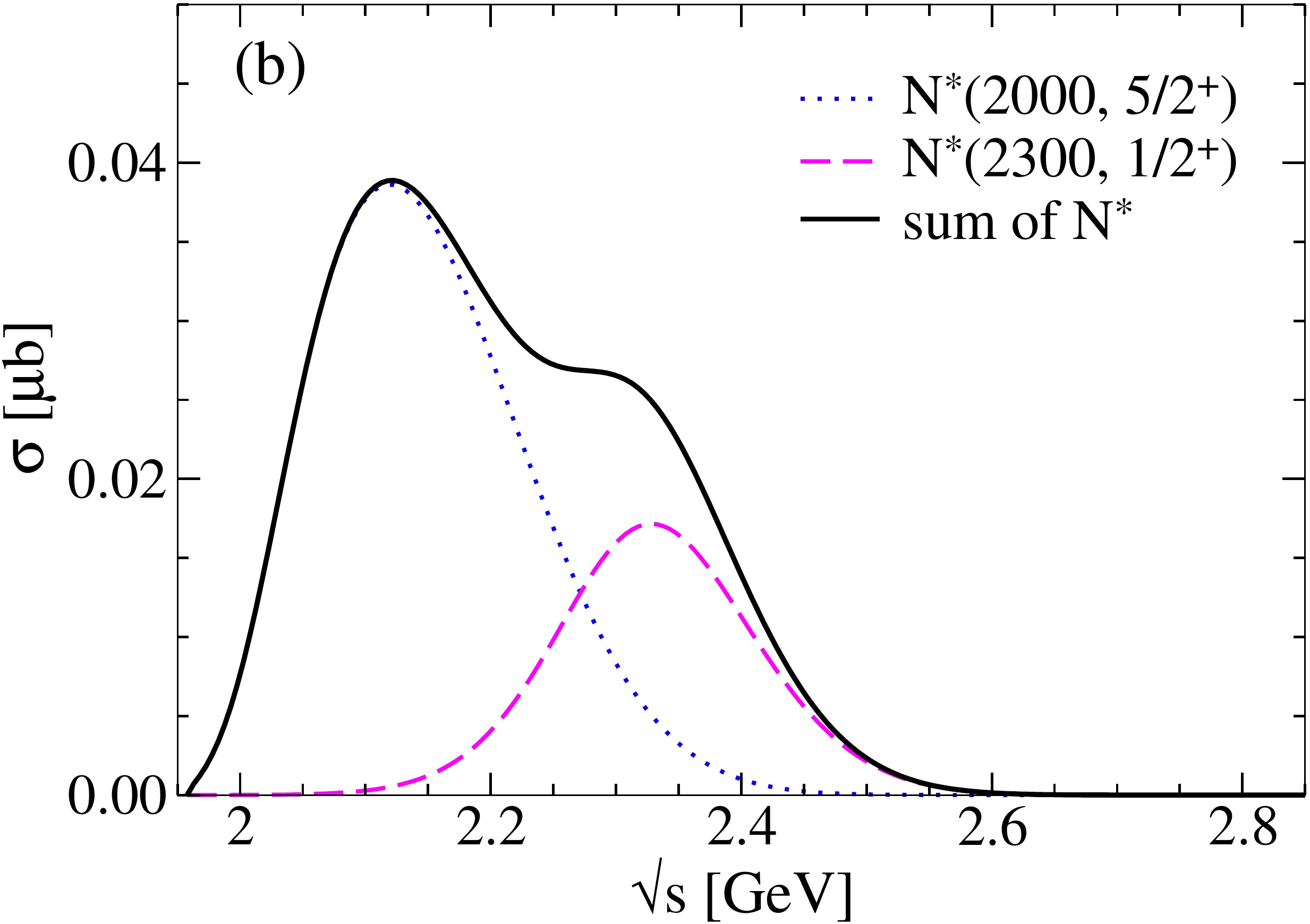}
\caption{(a) Total cross section is plotted as a function of the photon
laboratory energy $E_\gamma^{\mathrm{lab}}$.
The data are taken from Refs.~\cite{Ballam:1972eq,Barber:1981fj,Egloff:1979mg,
Busenitz:1989gq}.
(b) Each contribution of the two nucleon resonances as a function of the c.m.
energy $\sqrt{s}$.}
\label{fig:3}
\end{figure}
We are now in a position to present the numerical results and relevant
discussions.
We first fix the strength factor for the Pomeron exchange to be $C_{\mathbb{P}}=
6.5$ such that the formal $s \to \infty$ asymptotic behavior of the total cross
section is properly described.
Then the low-energy CLAS data~\cite{Seraydaryan:2013ija,Dey:2014tfa} are
used to constrain other model parameters:
The cutoff masses for PS- and S-meson exchanges are fixed to be $\Lambda_{\pi,\eta}
= 0.87$ and $\Lambda_{a_0,f_0} = 1.35$ GeV, respectively, and those for the
$\phi$-meson radiations and $N^*$ exchanges are fixed to be $\Lambda_{N,N^*} =
1.0$ GeV.
The unpolarized differential cross section can be expressed in terms of the
invariant amplitudes as follows:
\begin{align}
\label{eq:def:DCS}
\frac{d\sigma}{d\Omega} = \frac{1}{64\pi^2s} \frac{\bf |q|}{\bf |k|} \frac{1}{4}
\sum_{\mathrm{spins}} {|\sum_{h=B,N^*} \mathcal M_h (F_h)^n|}^2 ,
\end{align}
where $\bf k$ and $\bf q$ indicate the three-momenta of the incoming photon and
the outgoing $\phi$-meson, respectively, defined in the c.m. frame.
The exchanged particles are composed of the Born
($B$=$\mathbb{P}$,\,$f_1(1285)$,\,$\pi$,\,$\eta$,\,$a_0$,\,$f_0$,\,$N$) and
nucleon-resonance [$N^*$=$N^*(2000,5/2^+)$,\,$N^*(2300,1/2^+)$] terms.
As for the relevant phase factors between exchanged particles, the factor
$e^{i\pi/2}$ is additionally multiplied to the scalar exchange amplitude for a
better description of the experimental data.
We give the details of all the invariant amplitudes in Appendix~A. 

Figure~\ref{fig:3}(a) shows the total cross section for $\gamma p
\to \phi p$ as a function of the photon laboratory (lab) energy
$E_\gamma^{\mathrm{lab}}$.
Although the present theoretical setup is not applicable to the higher-energy
region far beyond the threshold, we plot the numerical results up to
$E_\gamma^{\mathrm{lab}} = 100$ GeV to see the tendency of the curves.
The conventional Pomeron exchange (green dotted curves) matches with the
intermediate-energy data at $E_\gamma^{\mathrm{lab}} = (3-10)$ GeV
~\cite{Ballam:1972eq,Barber:1981fj} but rather overestimates the data in the
high-energy range $E_\gamma^{\mathrm{lab}} \gtrsim 30$ GeV~\cite{Egloff:1979mg,
Busenitz:1989gq}.
The strength of the Pomeron exchange is almost the same as that of the Born
contribution.
AV-, PS-, and S-meson exchanges have small effects on the total cross section
but come into play significantly for the differential cross sections and
spin-density matrix elements (SDMEs) as will be seen later.
The $\pi$- and $a_0$-meson exchanges turn out to be more important than the
$\eta$- and $f_0$-meson ones, respectively.
The structure shown at $E_\gamma^{\mathrm{lab}} = (1.6-3.0)$ GeV comes from the
contribution of the two nucleon resonances.
Each and total $N^*$ contributions are shown in Fig.~\ref{fig:3}(b) as a
function of the c.m. energy $\sqrt{s}$.
$N^*(2100,5/2^+)$ and $N^*(2300,1/2^+)$ are responsible for the peaks around
$\sqrt{s} \approx 2.1$ and $2.3$ GeV, respectively.

\begin{figure}[htp]
\centering
\includegraphics[width=17cm]{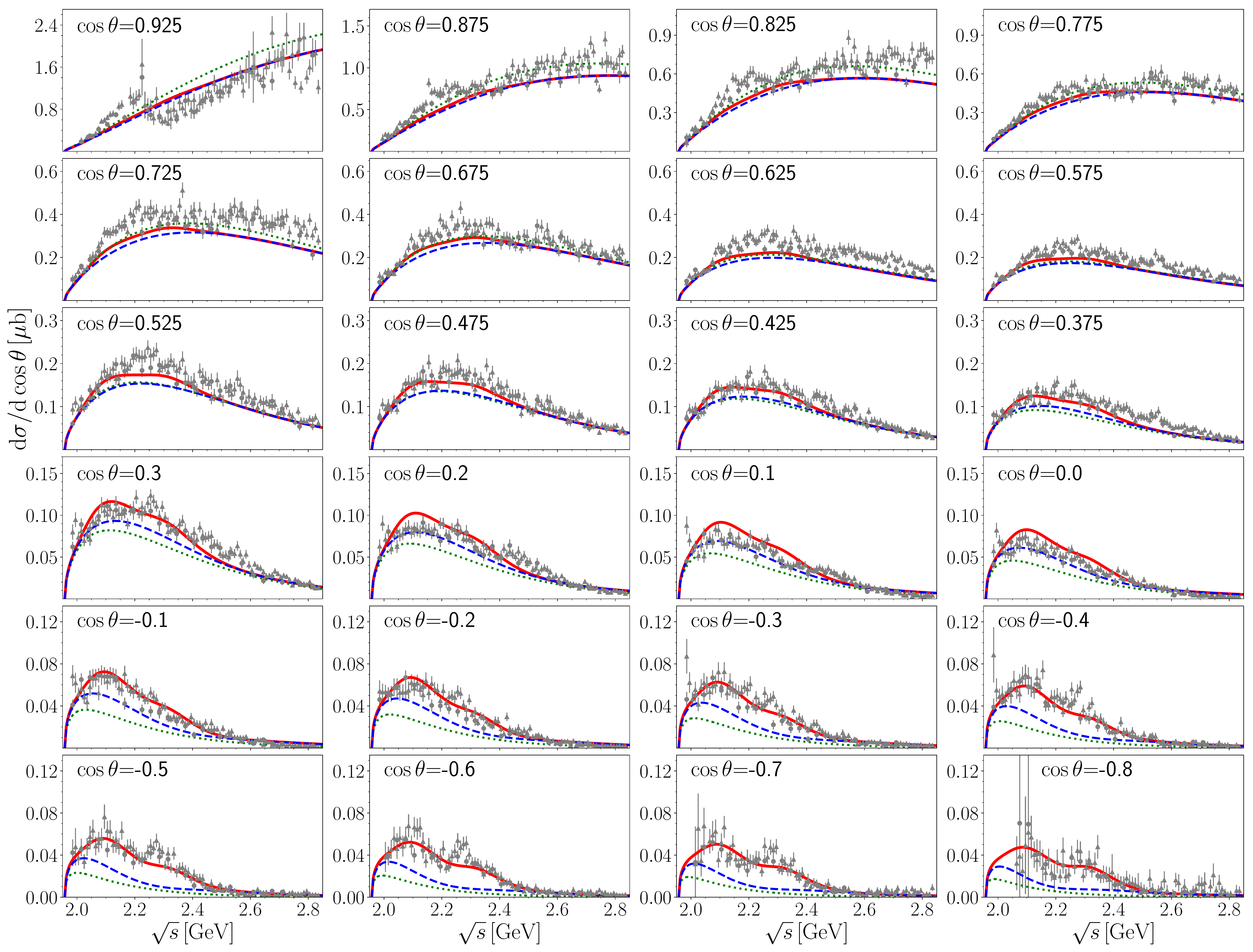}
\caption{Differential cross sections $d\sigma/d\cos\theta$ are plotted as
functions of the c.m. energy $\sqrt{s}$ at full scattering angles.
The red solid curves and the blue dashed curves stand for the total contribution
and the contribution without $N^*$, respectively. The green dotted curves stand
for the Pomeron contribution.
The CLAS data shwon by triangles and circles indicate the charged- and
neutral-$K \bar K$ decay modes of $\phi$, respectively~\cite{Dey:2014tfa}.}
\label{fig:4}
\end{figure}
We depict the differential cross sections as functions of $\sqrt{s}$ in
Fig.~\ref{fig:4} for wide scattering-angle regions in comparison to the charged
($\phi \to K^+ K^-$) and neutral ($\phi \to K_S^0 K_L^0$) decay modes
~\cite{Dey:2014tfa}, which are shown by the triangle and circle symbols,
respectively.
The angle $\theta$ is that of the outgoing $\phi$-meson in the c.m. frame.
Note that they are very similar to each other and should be treated as a single
data set and not be treated independently because they are related to each other
and the systematic uncertainties are greatly reduced~\cite{Dey:2014tfa}.
Except for the most forward angle $\cos\theta=0.925$, the total results turn out
to be reasonably successful over the wide scattering-angle regions.
The ratio of the Pomeron contribution to the CLAS data gradually decreases as
$\cos\theta$ decreases.
Note that PS- and S-meson exchanges, respectively, make constructive and
destructive interference effects with the dominant Pomeron exchange.
In the case of the SDMEs, the opposite interference pattern is observed.
As seen in Fig.~\ref{fig:3}, the contribution of the $a_0$-meson exchange is
about two times larger than that of the $\pi$-meson exchange.
Thus we expect the differential cross sections to be rather decreased with the
Pomeron plus $a_0$- and $\pi$-meson exchange model relative to the Pomeron
exchange one.
Thus we need to include one more ingredient in the $t$ channel and we select the
$f_1(1285)$ meson trajectory.
Its inclusion makes the differential cross sections enhanced to a certain extent.
That is why the Born contribution is almost the same as the Pomeron contribution
at the intermediate angles and even prevails over it at the backward ones.
Nevertheless, the ratio of the Born contribution to the CLAS data at the backward
angles $\cos\theta \lesssim -0.5$ is even less than 50\%.
Two $N^*$ contributions improve theoretical results remarkably, i.e., the two
peaks around $\sqrt{s} \approx 2.1$ and $2.3$ GeV can be accounted for by
the effects of $N^*(2000,5/2^+)$ and $N(2300,1/2^+)$, respectively.
The clear bump structure at $\cos\theta = 0.925$ around $\sqrt{s} \approx
2.2$ GeV, admittedly, may arise from another mechanism, the study of which is 
beyond the scope of the present work.

\begin{figure}[htp]
\centering
\includegraphics[width=17cm]{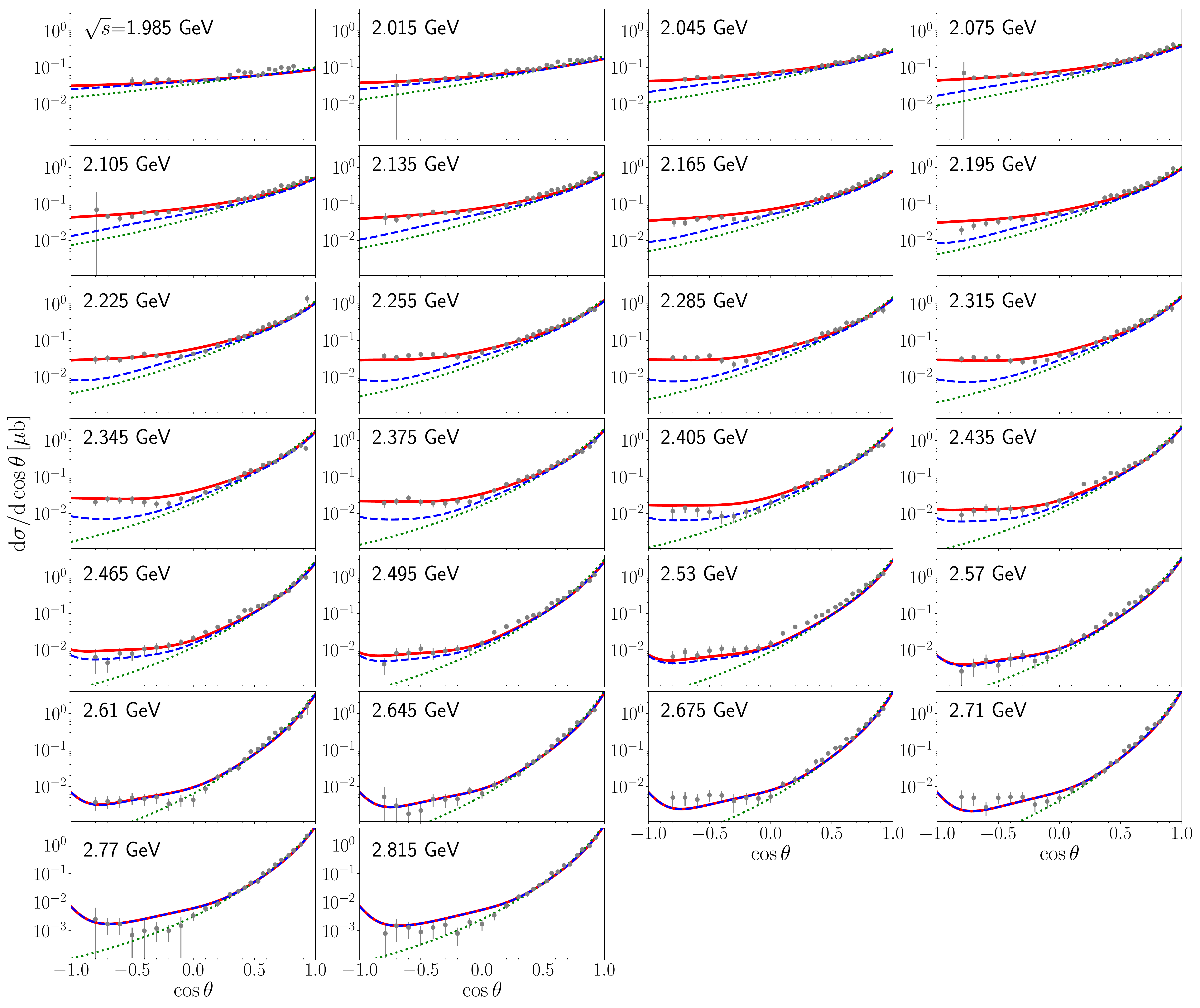}
\caption{Differential cross sections $d\sigma/d\cos\theta$
are plotted as functions of $\cos\theta$ for different c.m. energies
$\sqrt{s} = (1.985 - 2.815)$ GeV.
The curve notations are the same as those in Fig.~\ref{fig:4}.
The CLAS data indicate the neutral-$K \bar K$ decay mode of $\phi$
~\cite{Dey:2014tfa}.}
\label{fig:5}
\end{figure}
The differential cross sections $d\sigma/d\cos\theta$ are displayed in
Fig.~\ref{fig:5} as functions of $\cos\theta$ for different c.m. energy bins in
the logarithmic scale.
The Pomeron contribution (green dotted curves) governs the forward angle
regions, but starts to deviate from the CLAS data~\cite{Dey:2014tfa} as
$\cos\theta$ decreases.
The inclusion of AV-, PS-, and S-meson exchanges (blue dashed curves) makes the
curves increased in the region of $\cos\theta \lesssim 0.5$.
The remaining discrepancies in the range of $\sqrt{s} =(2.04-2.50)$ GeV are
reduced by the effects of the $N^*$ contributions.
Consequently, the total results (red solid curves) are in very good agreement
with the CLAS data.

\begin{figure}[htp]
\centering
\includegraphics[width=17cm]{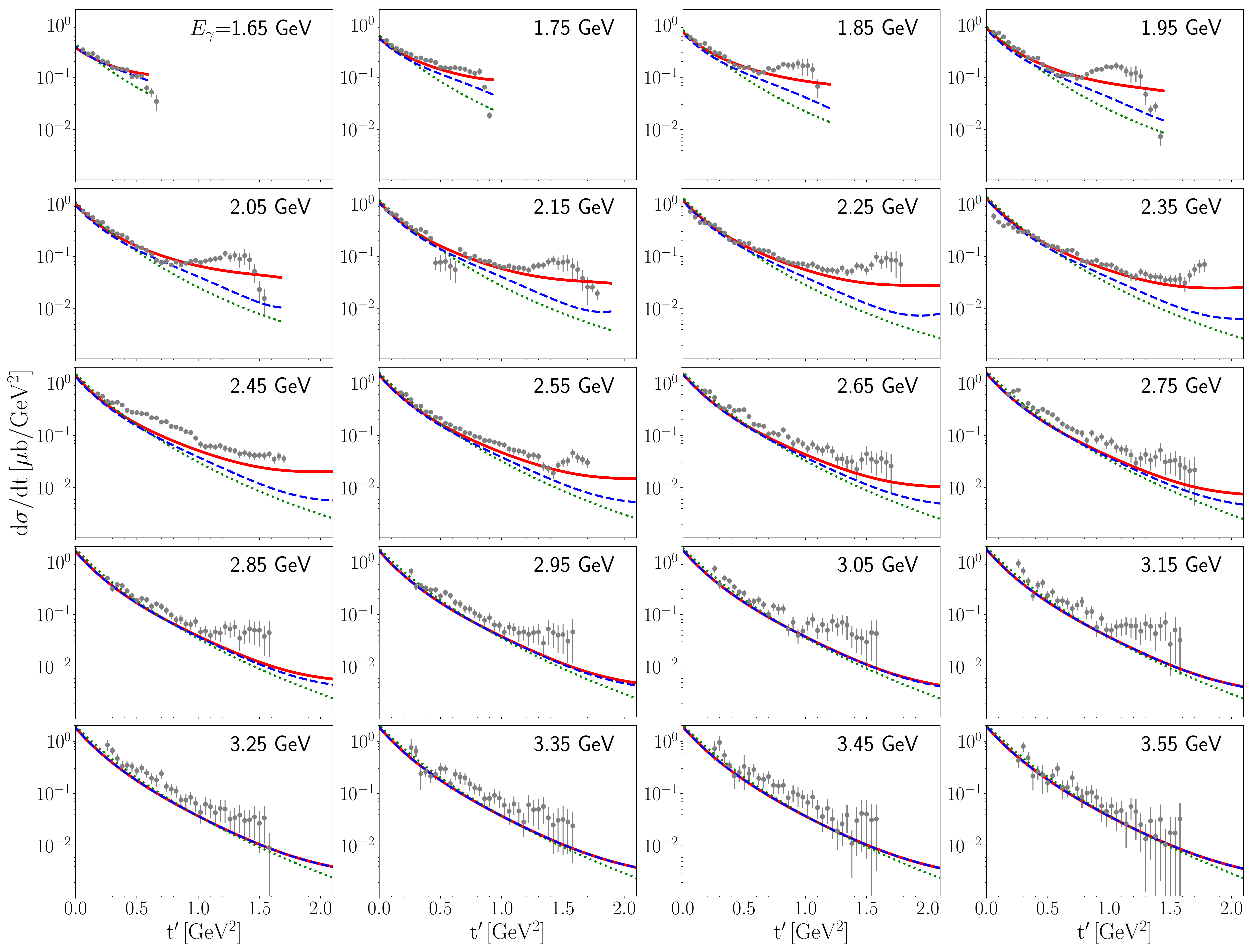}
\caption{Differential cross sections $d\sigma/dt$ are plotted as functions of
$t'\equiv|t-t_{\mathrm{min}}|$ for different laboratory energies $E_\gamma =
(1.65-3.55)$ GeV.
The curve notations are the same as those in Fig.~\ref{fig:4}.
The CLAS data indicate the neutral-$K \bar K$ decay mode of $\phi$
~\cite{Seraydaryan:2013ija}.}
\label{fig:6}
\end{figure}
In Fig.~\ref{fig:6}, we present the numerical results of the forward-scattering
cross sections $d\sigma/dt$ as functions of the momentum transfer squared
$t'\equiv|t-t_{\mathrm{min}}|$ for different laboratory energy bins where
$t_{\mathrm{min}}$ indicates the minimum value of $t$ at a certain fixed energy.
The tendency is similar to Fig.~\ref{fig:5} in general for the corresponding
beam energies.
The level of agreement between the total results and the CLAS data
~\cite{Seraydaryan:2013ija} is quite good at $E_\gamma^{\mathrm{lab}} \gtrsim 2.45$
GeV.
However, we find the bump structures at large values of $t'$ near the threshold
$E_\gamma^{\mathrm{lab}} = (1.85-2.35)$ GeV, indicating that the $N^*$
contributions should be treated with caution.

\begin{figure}[h]
\centering
\includegraphics[width=5.5cm]{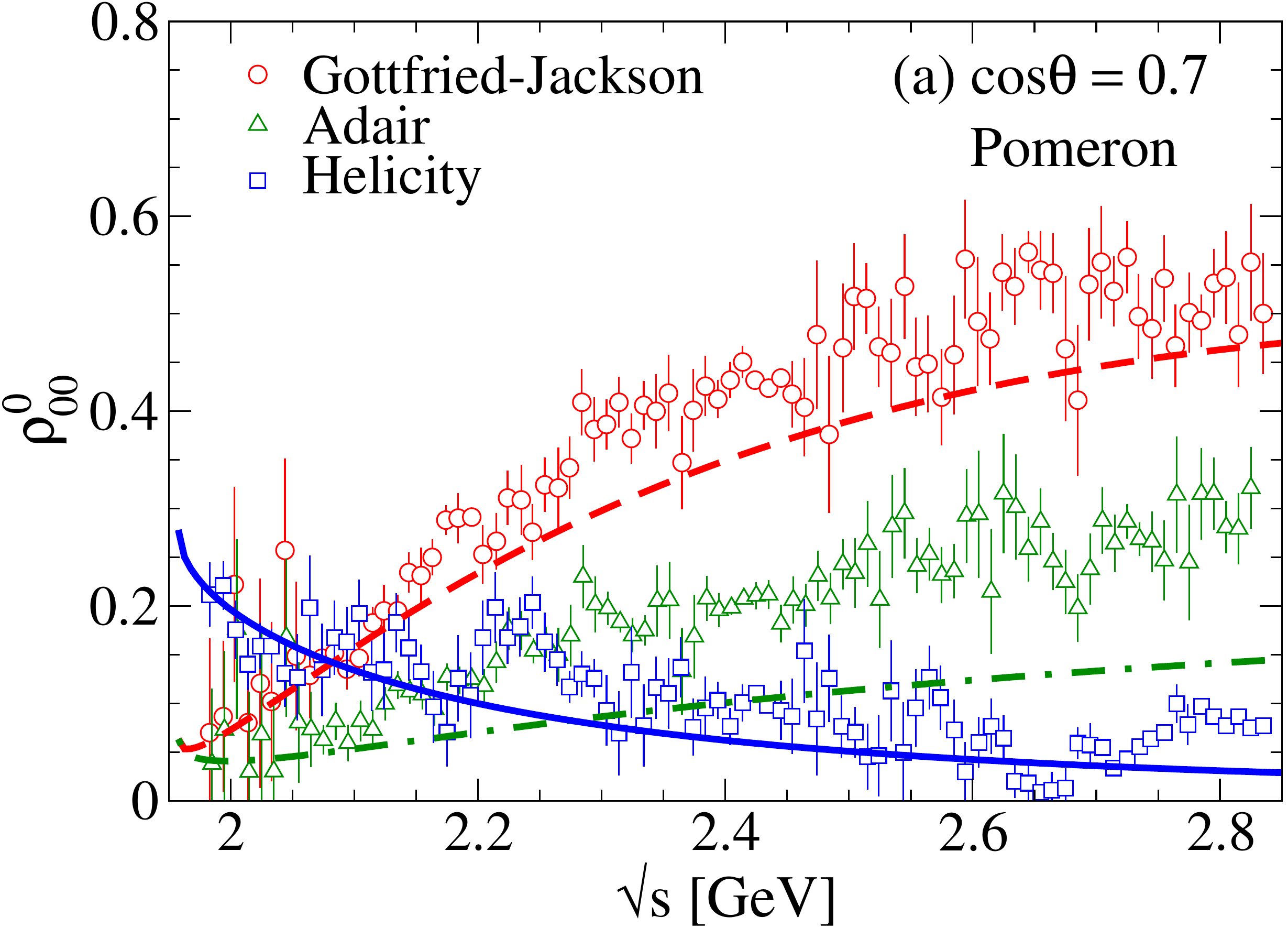} \hspace{1em}
\includegraphics[width=5.5cm]{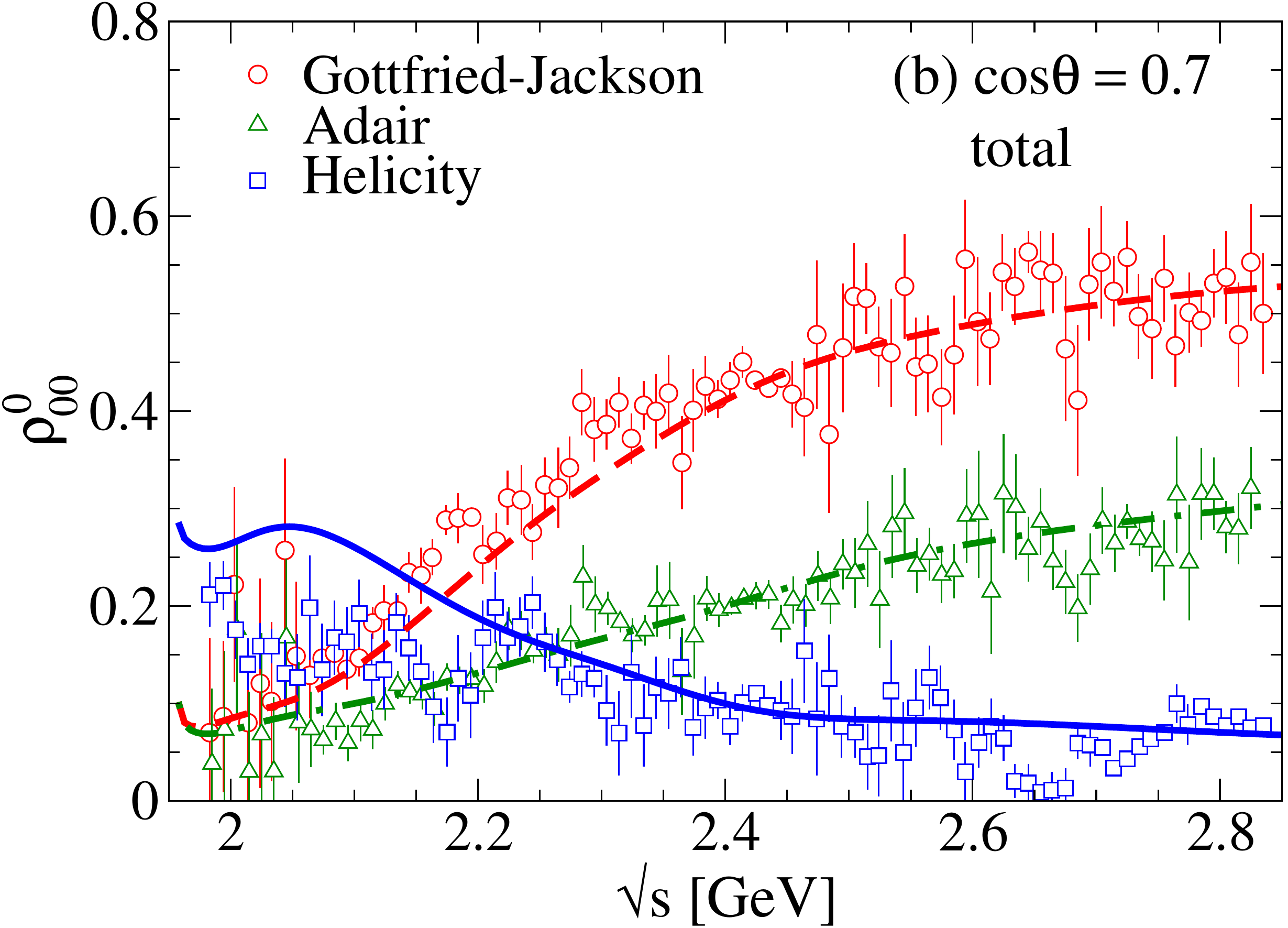}
\caption{Spin-density matrix elements $\rho^0_{00}$ are plotted as functions
of $\sqrt{s}$ at $\cos\theta = 0.7$ in the three different reference frames.
The red dotted, green dot-dashed, and blue solid curves indicate the results in
the Gottfried-Jackson, Adair, and helicity frames, respectively, which
correspond to the CLAS data (red circles, green triangles, and blue squares)
from the charged-$K \bar K$ decay mode of $\phi$~\cite{Dey:2014tfa}.
The results in panels (a) and (b) stand for the Pomeron and total
contributions, respectively.}
\label{fig:7}
\end{figure}
From now on, we present the results of SDMEs~\cite{Schilling:1969um} in
various reference frames to shed light on the relevant reaction mechanism.
Their definitions are given in Appendix~B.
Figure~\ref{fig:7} depicts them as a function of $\sqrt{s}$ at $\cos\theta =
0.7$ in the Gottfried-Jackson (red dashed curves), Adair (green dot-dashed
curves), and helicity (blue solid curves) frames~\cite{Dey:2014tfa}.
We observe definitely nonzero values of $\rho_{00}^0$ in the Gottfried-Jackson and
helicity frames, which show that $t$-channel (TCHC) and $s$-channel (SCHC)
helicity conservations are broken, respectively.
The diffractive Pomeron exchange is expected to be dominant at the
forward-scattering angles, but underestimates the $\rho_{00}^0$ data in all three
frames as depicted in Fig.~\ref{fig:7}(a).
The finite values of $\rho_{00}^0$ reflect the single helicity-flip transition
between the incoming photon and the outgoing $\phi$ meson from its definition as
understood by
\begin{align}
\label{eq:def:Rho00}
\rho_{00}^0 \propto |\mathcal{M}_{\lambda_{\gamma = 1},\lambda_{\phi = 0}}|^2 +
|\mathcal{M}_{\lambda_{\gamma = -1},\lambda_{\phi = 0}}|^2 .
\end{align}
The Pomeron exchange is known as a gluon-rich Regge trajectory with a vacuum
quantum number ($J^{PC} = 0^{++}$) and thus we expect TCHC in principle.
Moreover, the argument in support of SCHC for diffractive vector meson
photoproductions is given in the literature~\cite{Donnachie:1987abc,
Gilman:1970vi}.
However, there is no clear reason why TCHC and SCHC should hold for our
phenomenological DL model~\cite{Donnachie:1987abc}.
The total results finally succeed in a satisfactory description of $\rho_{00}^0$
in all three frames as displayed in the Fig.~\ref{fig:7}(b).
Thus, the relative contributions of AV-, PS-, and S-meson exchanges to the
dominant Pomeron contribution are confirmed more explicitly.
We can immediately test a simple DL Pomeron plus $\pi$- and $\eta$-meson exchange
model~\cite{Titov:1999eu} via the present SDME data.
It turns out that the agreement between these model predictions and the SDME
data is not satisfactory in all three frames.
The comparison with the LEPS data also supports this argument as shown in Fig.~6
of Ref.~\cite{Chang:2010dg}.

\begin{figure}[htp]
\centering
\includegraphics[width=15cm]{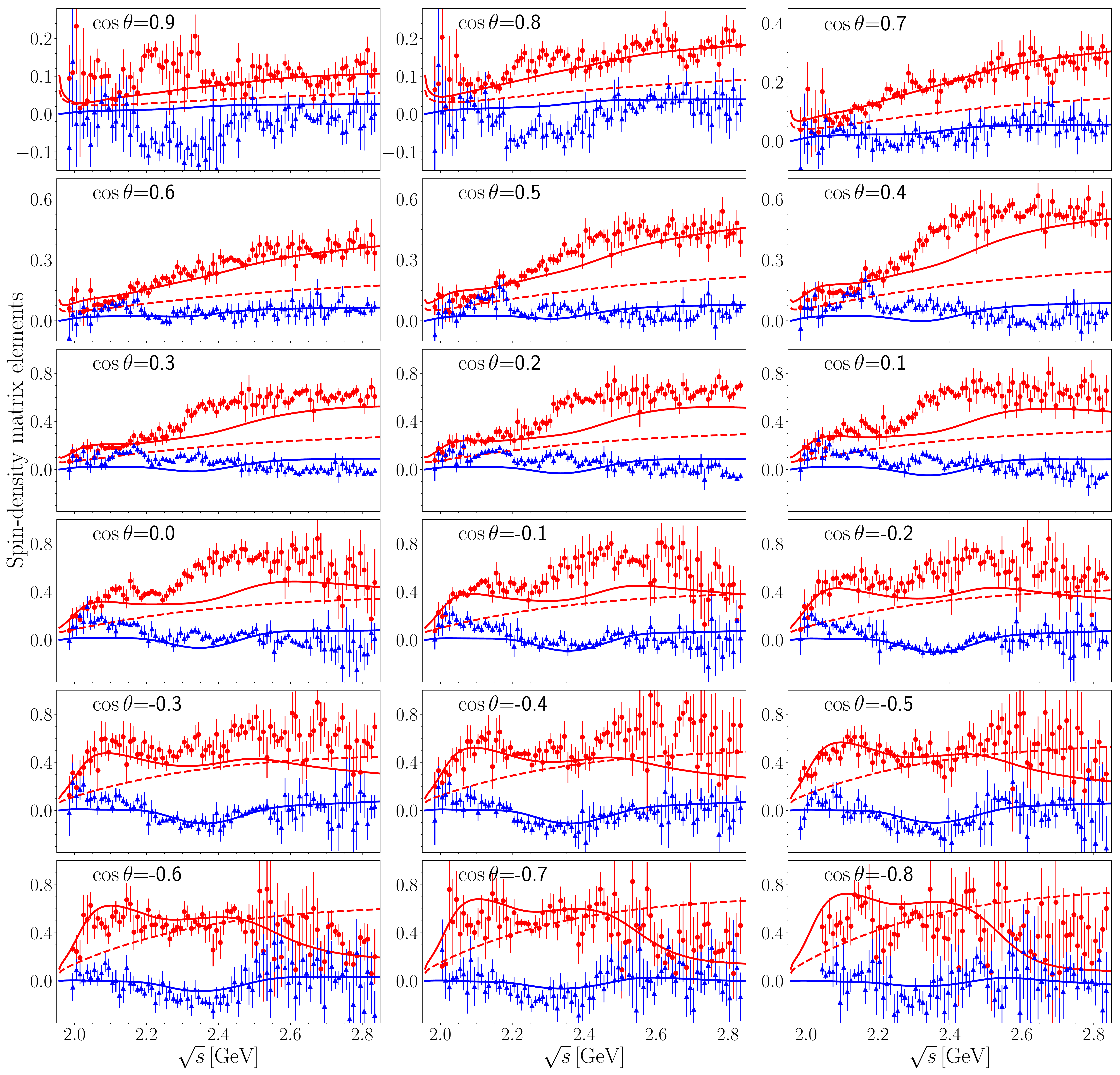}
\caption{Spin-density matrix elements $\rho_{00}^0$ (red circles) and
$\rho_{1-1}^0$ (blue triangles) are plotted as functions of $\sqrt{s}$ at full
scattering angles in the Adair frame.
The red dashed curve indicates the results of $\rho_{00}^0$ for Pomeron
exchange, and red solid curves and blue solid curves correspond to the total
results of $\rho_{00}^0$ and $\rho_{1-1}^0$, respectively.
The CLAS data indicate the charged-$K \bar K$ decay mode of
$\phi$~\cite{Dey:2014tfa}.}
\label{fig:8}
\end{figure}
It is worthwhile to examine other components of SDMEs to understand the effects
of the various contributions.
Figure~\ref{fig:8} depicts $\rho_{00}^0$ (red circles) and $\rho_{1-1}^0$
(blue triangles) as functions of $\sqrt{s}$ at full scattering angles in the
Adair frame~\cite{Dey:2014tfa}.
It is noticeable that the $\rho_{00}^0$ data are quite large, unlike the
$\rho_{1-1}^0$ data which are small but finite.
The $\rho_{00}^0$ data are all positive and reveal bumplike structures at the
threshold of $\sqrt{s}=(2.0-2.2)$ GeV and backward-scattering angles of
$\cos\theta \lesssim 0.2$, even though systematical limitations at the angles
make the structures unclear.
First, we find that the Pomeron exchange alone (red dashed curves) is not
sufficient for describing the $\rho_{00}^0$ data at the forward-scattering angles.
The inclusion of the S mesons to the Pomeron exchange makes $\rho_{00}^0$
increased, but the inclusion of the PS mesons makes the results worse by pulling
down $\rho_{00}^0$.
The cutoff masses in the form factor of Eq.~(\ref{eq:FF:M}) are
constrained so as to describe simultaneously the differential cross sections and
SDMEs.
Second, it turns out that the bumplike structures observed at the
backward-scattering angles is a clear evidence of the $N^*$ contribution.
At $\sqrt{s}=(2.0-2.2)$ GeV, both the $\rho_{00}^0$ data and the total theoretical
results (red solid curves) are the strongest at very backward angles, and get
reduced steadily with respect to $\cos\theta$, and then vanish around
$\cos\theta = 0.2$.
This tendency is almost the same as the numerical results of the differential
cross sections in Fig.~\ref{fig:4}.
The Born contribution merely underestimates the $\rho_{00}^0$ data for all the
available energies.
The $N^*(2000,5/2^+)$ contribution is mainly responsible for the local structures
rather than other spin-$1/2$ and-$3/2$ nucleon resonances.
The total results are in good agreement with the CLAS data in general except in
the intermediate-angle region where the interference between the Born and
resonant terms is maximal.

\begin{figure}[htp]
\centering
\includegraphics[width=8.3cm]{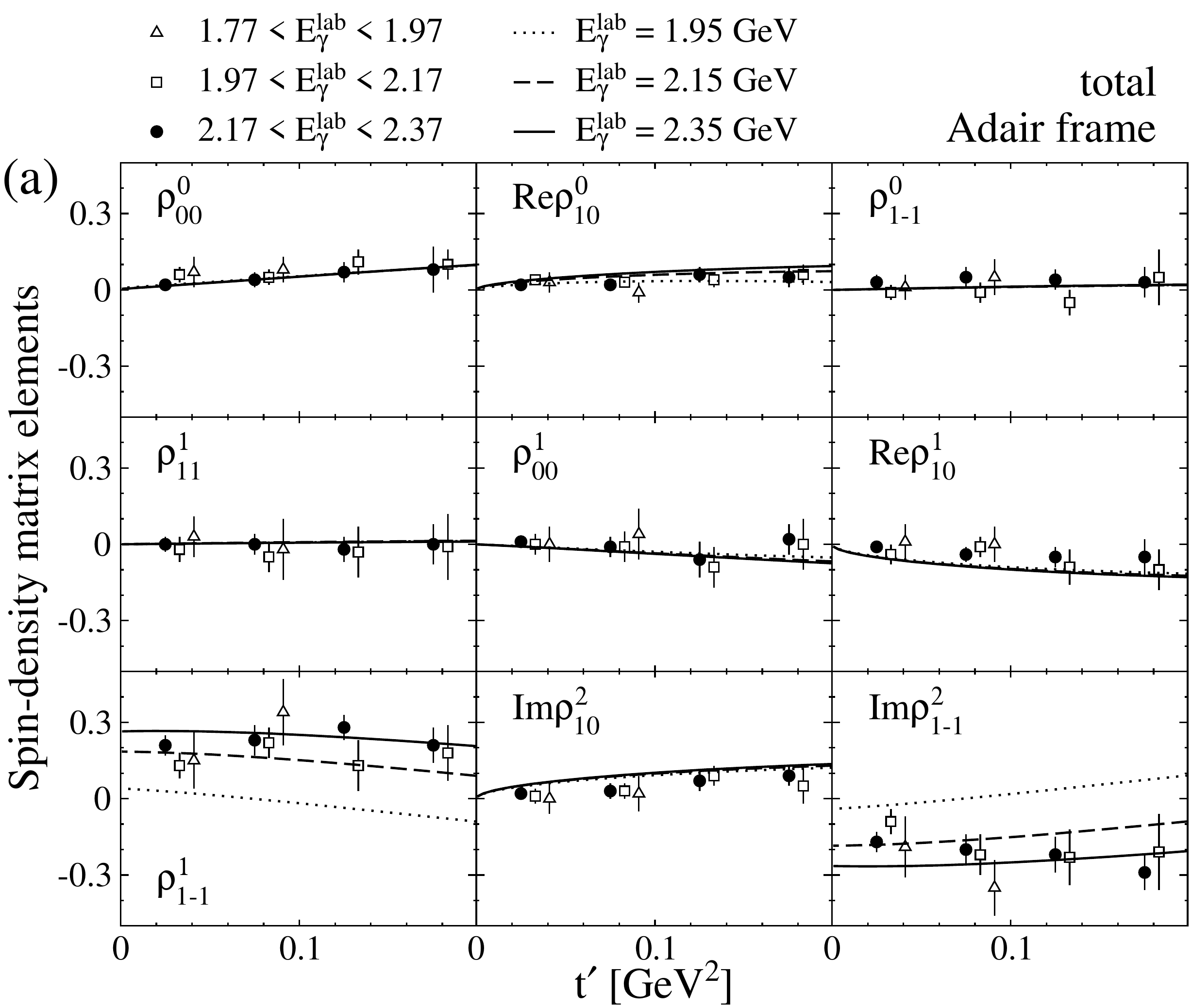} \,\,\,\,\,
\includegraphics[width=8.3cm]{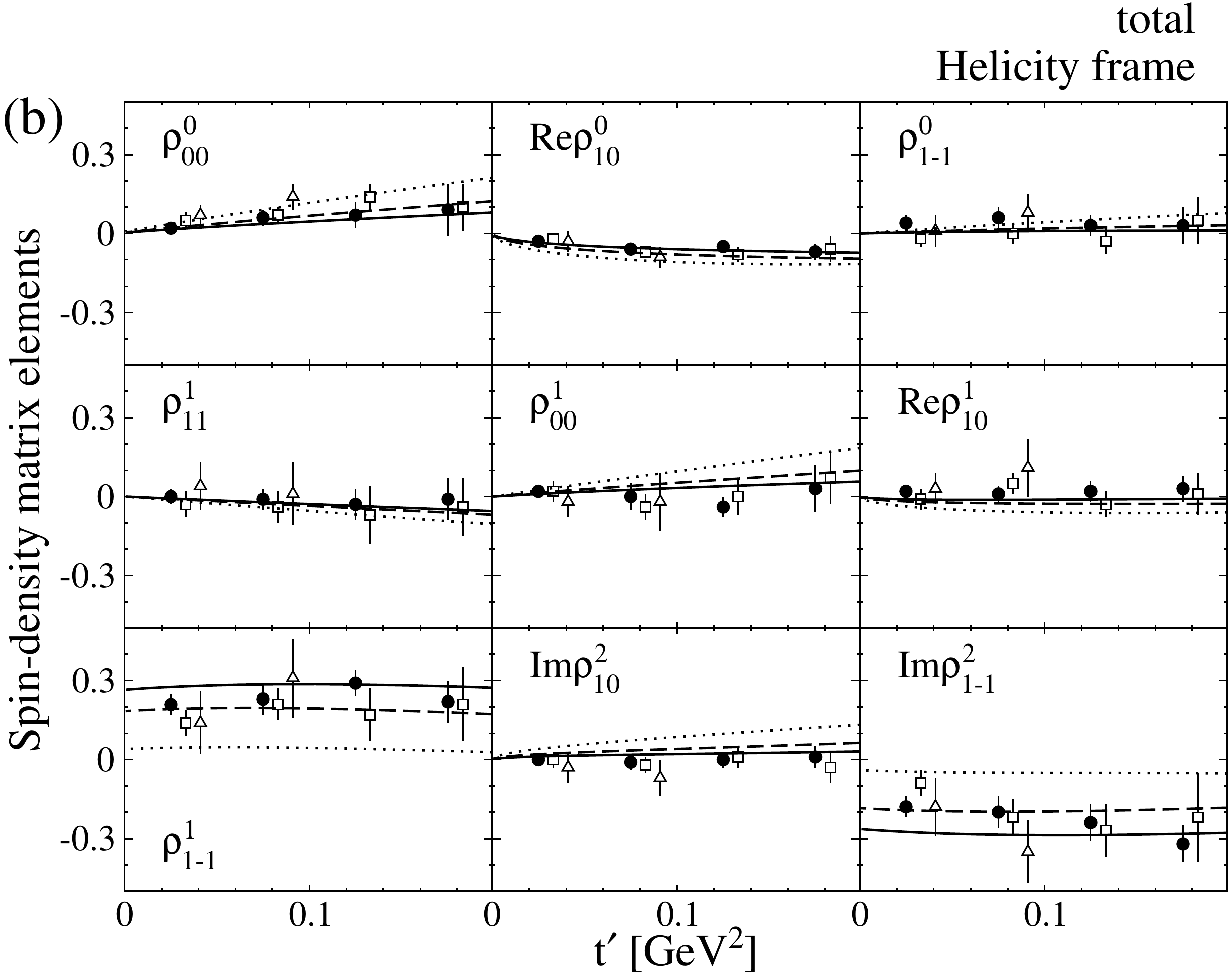}
\caption{The total contribution of various spin-density matrix elements are
plotted as functions of $t'\equiv |t-t_{\mathrm{min}}|$ for three different
laboratory energies, $E^\mathrm{lab}_\gamma= 1.95$ GeV (dotted curves), $2.15$
GeV (dashed curves), and $2.35$ GeV (solid curves), in the Adair (a) and
helicity (b) frames.
The data are from the LEPS Collaboration~\cite{Chang:2010dg}.}
\label{fig:9}
\end{figure}
\begin{figure}[htp]
\centering
\includegraphics[width=8.5cm]{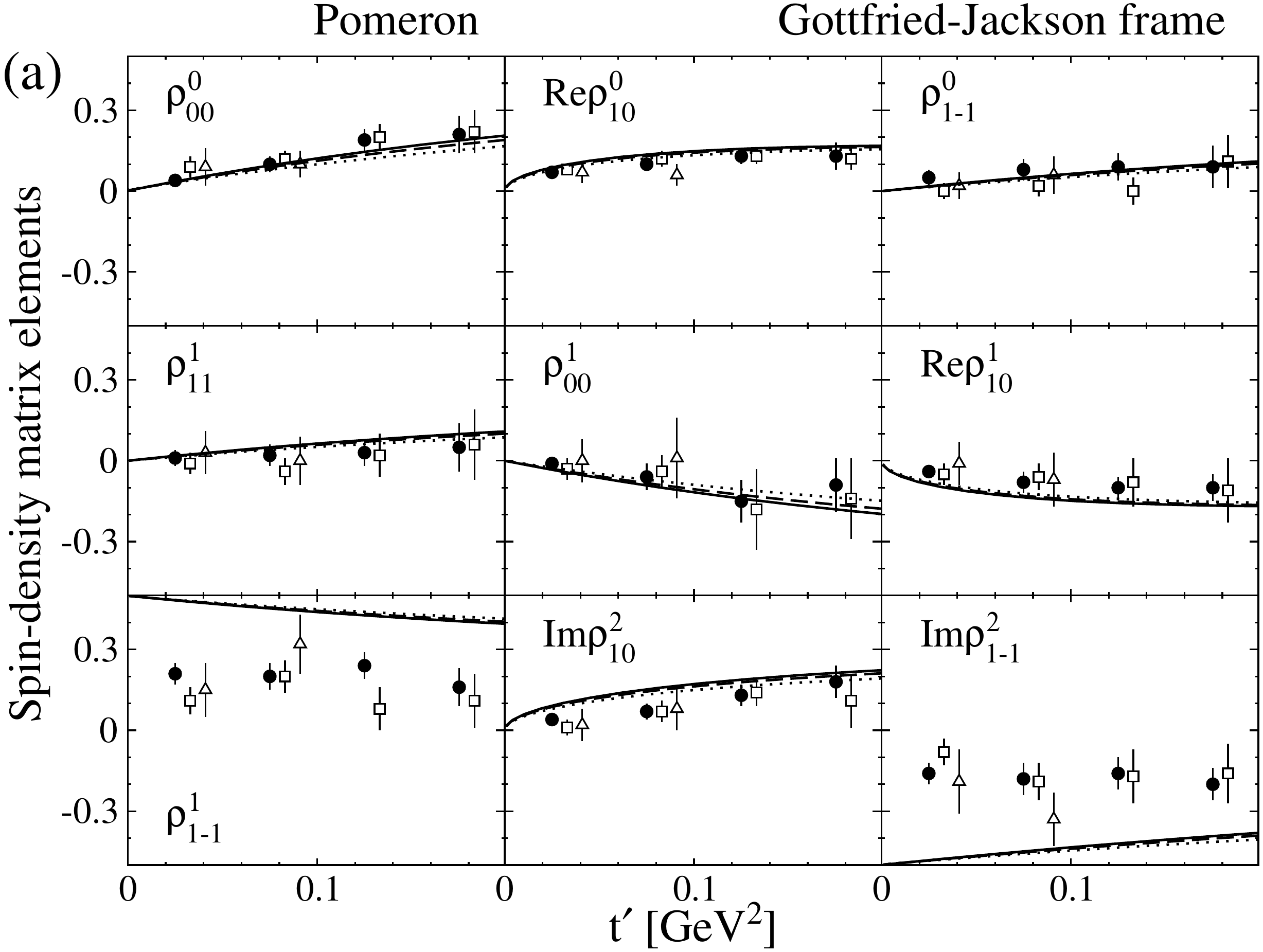} \,\,\,\,\,
\includegraphics[width=8.5cm]{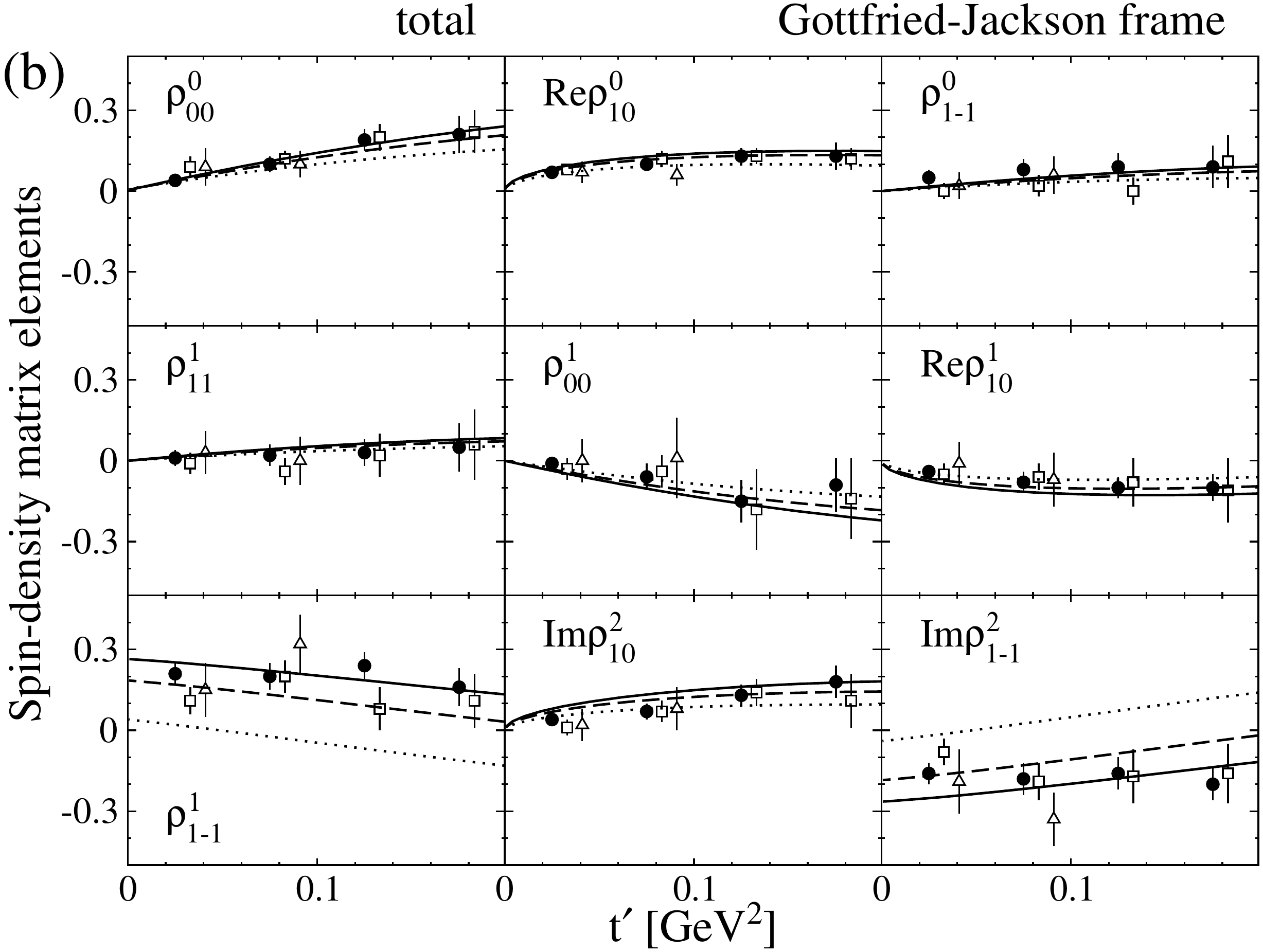}
\caption{The same as Fig.~\ref{fig:9} but for the Pomeron (a) and total (b)
contributions, respectively, in the Gottfried-Jackson frame.}
\label{fig:10}
\end{figure}
Let us continue to present our results in comparison to the LEPS
data~\cite{Chang:2010dg}.
The total contribution of various SDMEs are displayed in Figs.~\ref{fig:9}(a)
and \ref{fig:9}(b) as functions of $t'\equiv|t-t_{\mathrm{min}}|$ in the Adair and
helicity frames, respectively.
Here we examine three different threshold energies, i.e., $E^\mathrm{lab}_\gamma=
1.95$ GeV (dotted curves), $2.15$ GeV (dashed curves), and $2.35$ GeV (solid
curves).
We find distinctive large values for $\rho_{1-1}^1$ and
$\mathrm{Im}[\rho_{1-1}^2]$.
Their similar absolute magnitudes are understood from the following
relation~\cite{Titov:2003bk},
\begin{align}
\label{eq:RhoRela}
- \mathrm{Im}[\rho_{1-1}^2] \approx \rho_{1-1}^1 -
\frac{(\rho_{1-1}^0)^2}{1-\rho_{00}^0} ,
\end{align}
because small values of $\rho_{1-1}^0$ are experimentally observed.
The data other than $\rho_{1-1}^1$ and $\mathrm{Im}[\rho_{1-1}^2]$ are almost 
zero and are in good agreement with our results.
The considered region $t'=(0.0-0.2)\,\mathrm{GeV^2}$ is dominated by the
$t$-channel exchange process for the differential cross sections but is sensitive
to the structure of the $N^*$ exchange amplitudes for the case of SDMEs.

It is more informative to present our results in the Gottfried-Jackson
frame for which we separately show the Pomeron and total contributions in
Figs.~\ref{fig:10}(a) and~\ref{fig:10}(b), respectively.
The increase in the magnitude of $\rho_{00}^0$ is consistent with the results
of Fig.~\ref{fig:8}. When a double helicity-flip transition is forbidden, the
$\rho_{1-1}^0$ is exactly zero by construction.
However, the small but finite value of $\rho_{1-1}^0$ at large $t'$ even for
Pomeron exchange implies the possible spin-orbital interaction from our modified
DL model~\cite{Titov:1999eu,Laget:2000gj,Chang:2010dg}.
For the pure Pomeron exchange process, the following relation
~\cite{Titov:2003bk},
\begin{align}
\label{eq:Rho1-1Rho00}
\rho_{1-1}^1 \simeq \frac{1}{2} (1 -\rho_{00}^0) ,
\end{align}
also holds and is confirmed by our numerical results.
Note that $\rho_{1-1}^1$ is very interesting, because it allows us to measure the
asymmetry of the relative strength between the natural ($N$) and unnatural ($U$)
parity exchange processes, and is written as
\begin{align}
\label{eq:def:Rho1-1}
\rho_{1-1}^1 = \frac12 \frac{\sigma^N - \sigma^U}{\sigma^N + \sigma^U} +
\frac12 \rho_{00}^1 .
\end{align}
For example, PS- and S-meson exchanges correspond to unnatural and natural parity
exchanges, respectively, and no single helicity-flip transition occurs, resulting
in $\rho_{00}^1 = 0$.
Thus we exactly obtain $\rho_{1-1}^1  = -0.5$ and $0.5$ for them, respectively.
A large deviation of $\rho_{1-1}^1$ from Pomeron exchange is compensated by the
inclusion of AV- and PS-meson unnatural parity exchanges and the $N^*$
contribution which involves both natural and unnatural parity exchanges.
The direct $\phi$-meson radiation hardly affects the SDMEs or the cross
sections.
The total results are improved in comparison to the results of Pomeron exchange.

\begin{figure}[h]
\centering
\includegraphics[width=9.8cm]{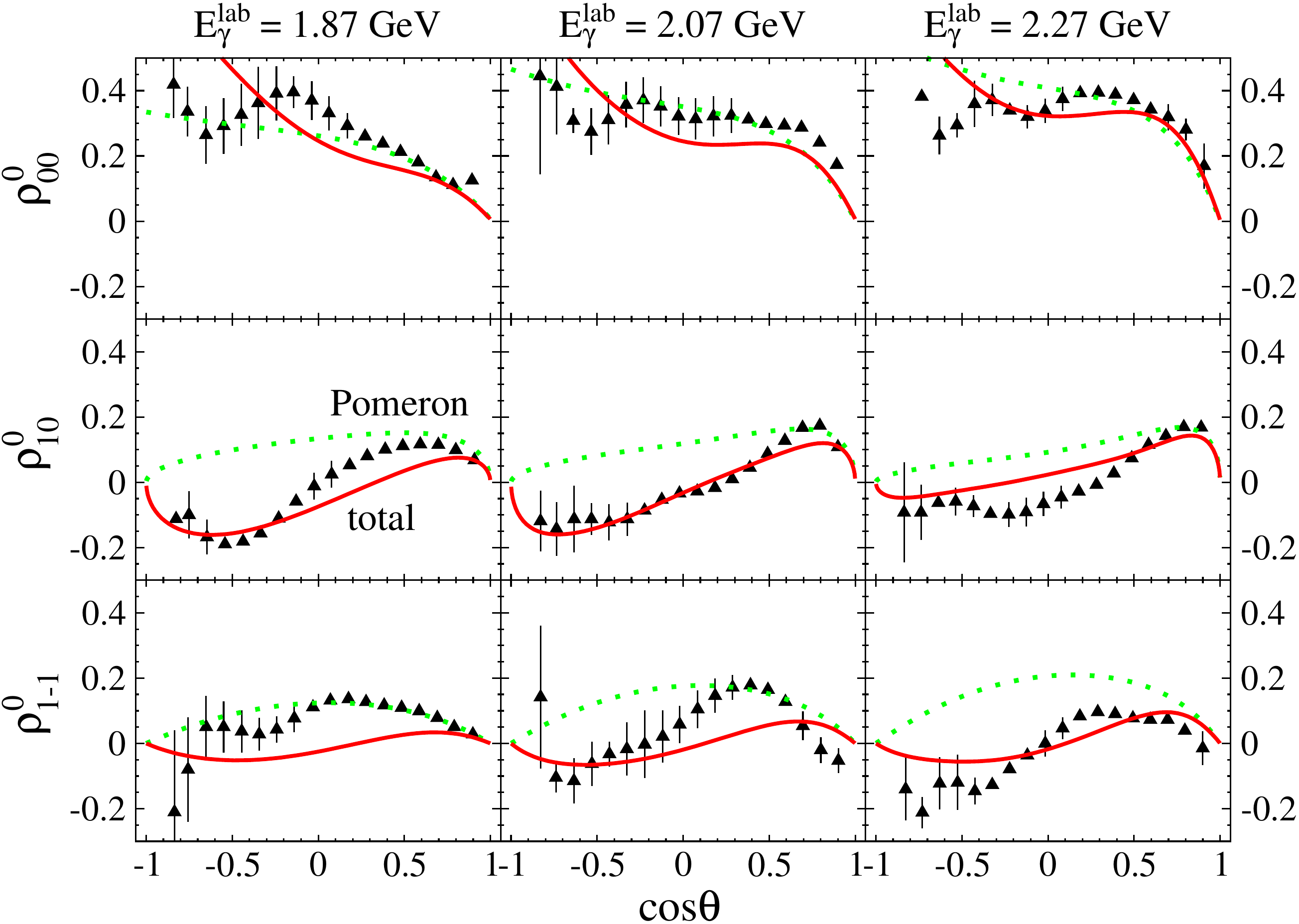}
\caption{Spin-density matrix elements $\rho_{00}^0$, $\rho_{10}^0$, and
$\rho_{1-1}^0$ are plotted as functions of $\cos\theta$ for three different
lab. energies in the Gottfried-Jackson frame.
The green dotted curves and the red solid curves stand for the Pomeron and total
contributions, respectively.
The CLAS data indicate the charged-$K \bar K$ decay mode of
$\phi$~\cite{Dey:2014tfa}.}
\label{fig:11}
\end{figure}
Figure~\ref{fig:11} shows the results of $\rho_{00}^0$, $\rho_{10}^0$, and
$\rho_{1-1}^0$ as functions of $\cos\theta$ in the Gottfried-Jackson frame for
three different threshold energies as done in Figs.~\ref{fig:9} and \ref{fig:10}.
The CLAS energy bins are 10 MeV wide~\cite{Dey:2014tfa}, but the LEPS ones
much wider, i.e., 200 MeV wide~\cite{Chang:2010dg}.
As expected, the forward-scattering angles are well described by Pomeron
exchange.
Large discrepancies shown at backward-scattering angles are highly decreased by
the inclusion of nucleon resonances, especially for $\rho_{10}^0$ and
$\rho_{1-1}^0$.

\section{Summary}
\label{SecIV}
We have investigated the $\phi$-meson photoproduction mechanism for the
wide-scattering-angle regions, based on the effective Lagrangian approach in the
tree-level Born approximation.
As for the Born contribution, we considered the universal Pomeron exchange and
the $f_1(1285)$ axial-vector-meson, ($\pi,\eta$) pseudoscalar-meson, and
($a_0,f_0$) scalar-meson exchanges, in addition to the direct $\phi$-meson
radiation contributions.
We newly took two nucleon resonances from the PDG into account, i.e.,
$N^*(2000, 5/2^+)$ and $N^*(2300,1/2^+)$.
We list important observations as follows.

\begin{itemize}
\item Pomeron exchange is responsible for the description of the available total
cross section data in the intermediate-energy, $E_\gamma^{\mathrm{lab}} = (3-10)$
GeV, and high-energy, $E_\gamma^{\mathrm{lab}} \gtrsim 30$ GeV, ranges.
\item The key point is how to incorporate the AV-, PS-, and S-meson exchanges
into the Pomeron contribution to describe the abundant differential
cross sections and SDMEs in the low energy region $\sqrt{s}= (2.0-2.8)$ GeV.
The $N^*$ contributions are also essential to account for the two
bumplike structures near $\sqrt{s} \approx 2.1$ and $2.3$ GeV shown at
the backward-scattering angles.
\item According to the interference patterns between the Born and $N^*$
contributions, the CLAS and LEPS data are reproduced qualitatively well, except
for the \textit{dip} or the \textit{peak} structure shown near the threshold in
the forward-scattering cross section.
We leave the long-standing puzzle on this structure to future work.
The role of the PS-meson exchanges is found to be definitely different from
that of the S-meson exchanges, and both are essential to describe the
cross sections and SDMEs simultaneously together with the AV-meson exchange.
\item The $N^*$ contributions should be dealt with carefully because of the lack
of the information on the $N^* \to \phi N$ decay channels.
We select $N^*(2000,5/2^+)$ and $N^*(2300,1/2^+)$ from the
PDG and extract their branching ratios by the comparison with the CLAS data.
We find that the bumplike structures observed in the backward-scattering angles
are explained qualitatively well by the $N^*$ contributions. 
\item We want to mention that the effect of high-spin meson exchanges
such as the $f_2'(1525)$ tensor-meson~\cite{Laget:2000gj} is revealed
especially on $\rho_{00}^0$ at forward-scattering angles to some extent.
However, we do not show these results because we want to emphasize that the
role of AV-, PS-, and S-meson exchanges is sufficient for the purpose of this
work.
\end{itemize}

Consequently, as shown in the present work, our rigorous theoretical analyses on
the presently available high-statistics and wide-angle-coverage experimental
data for the various physical observables will be valuable for a profound
understanding for the light-flavor ($\rho$ and $\omega$) and hidden-flavor
($J/\psi$) vector meson photoproductions~\cite{Mizutani:2017wpg,
Battaglieri:2001xv,Wu:2005wf,Williams:2009ab,Williams:2009aa,Vegna:2013ccx,
Wilson:2015uoa,Collins:2017vev}. Related works will appear elsewhere.

\section*{ACKNOWLEDGMENTS}
S.i.N. thanks B.~Dey, who kindly provided him with the experimental data for the
present work. 
The authors are grateful to A.~Hosaka (RCNP, Osaka) and A.~I.~Titov (JINR,
Dubna) for fruitful discussions.
This work was supported by the National Research Foundation of Korea (NRF)
(Grant No.~2018R1A5A1025563).
The work of S.i.N. was supported in part by a NRF grant
(Grant No.~2019R1A2C1005697).
The work of S.-H.Kim was supported in part by a NRF grant
(Grant No.~2019R1C1C1005790).
\section*{Appendix A: Invariant  Amplitudes}
\label{AppenA}
The invariant amplitude for $\gamma p \to \phi p$ can be written as follows
\begin{align}
\label{eq:AmpDef}
\mathcal{M} =
\varepsilon_\nu^* \bar{u}_{N'} \mathcal{M}^{\mu\nu} u_N \epsilon_\mu,
\end{align}
where $\epsilon_\mu$ and $\varepsilon_\nu$ stand for the polarization vectors
for the photon and the $\phi$ meson, respectively.
The Dirac spinors of the incident and outgoing nucleons are denoted by $u_N$
and $u_{N'}$, respectively.

The invariant amplitude for the Pomeron exchange is given by
\begin{align}
\label{eq:AmpPom}
\mathcal{M}_{\mathbb P}^{\mu\nu} = - M_{\mathbb P}(s,t) \Gamma_{\mathbb P}^{\mu\nu},
\end{align}
where the scalar function is
\begin{align}
\label{eq:ScaFun}
M_{\mathbb P}(s,t) = C_{\mathbb P} F_\phi(t) F_N(t) \frac{1}{s}
\left( \frac{s}{s_{\mathbb P}} \right)^{\alpha_{\mathbb P}(t)}
\mathrm{exp} \left[ -\frac{i\pi}{2}\alpha_{\mathbb P}(t) \right],
\end{align}
and the transition operator is
\begin{align}
\label{eq:TranOp}
\Gamma_{\mathbb P}^{\mu\nu} =
\left( g^{\mu\nu} - \frac{k_2^\mu k_2^\nu}{k_2^2} \right) \rlap{\,/}{k_1} -
\left( k_1^\nu - \frac{k_2^\nu k_1 \cdot k_2}{k_2^2} \right) \gamma^\mu -
\left( \gamma^\nu - \frac{\rlap{\,/}{k_2}k_2^\nu}{k_2^2} \right) k_2^\mu.
\end{align}
The energy-scale factors for the Pomeron is given by $s_{\mathbb P}=(M_N+M_\phi)^2$.
Practically, a phenomenological consideration gives a change for the last
term of Eq.~(\ref{eq:TranOp})
\begin{align}
\label{eq:k2Mod}
k_2^\mu \to k_2^\mu - \frac{(p_1+p_2)^\mu k_1 \cdot k_2}{(p_1+p_2)
\cdot k_1},
\end{align}
to satisfy the Ward-Takahashi identity in $\bar u_{N'} [\cdots] u_{N}$
of Eq.~(\ref{eq:AmpDef})~\cite{Titov:2003bk}.

The invariant amplitudes for $f_1(1285)$ axial-vector-, ($\pi,\eta$)
pseudoscalar-, and ($a_0,f_0$) scalar-meson exchanges take the following forms:
\begin{align}
\label{eq:Amp:M}
\mathcal{M}_{f_1}^{\mu\nu} = i \frac{M_\phi^2 g_{\gamma \phi f_1}g_{f_1 NN}}{t-M_{f_1}^2}
\epsilon^{\mu\nu\alpha\beta}
\left[ -g_{\alpha\lambda}+\frac{q_{t\alpha} q_{t\lambda}}{M_{f_1}^2} \right]
\left[ \gamma^\lambda + \frac{\kappa_{f_1 NN}}{2M_N} \gamma^\sigma \gamma^\lambda
q_{t\sigma} \right] \gamma_5 k_{1 \beta} ,
\end{align}
\begin{align}
\label{eq:Amp:PSS}
\mathcal{M}_\Phi^{\mu\nu} =& i\frac{e}{M_\phi}
\frac{g_{\gamma\Phi\phi}g_{\Phi NN}}{t-M_\Phi^2} \epsilon^{\mu\nu\alpha\beta} k_{1\alpha}
k_{2\beta} \gamma_5 ,
\cr
\mathcal{M}_S^{\mu\nu} =& \frac{e}{M_\phi}
\frac{2g_{\gamma S \phi}g_{S NN}}{t- M_S^2 + i\Gamma_S M_S}
(k_1 \cdot k_2 g^{\mu\nu} - k_1^\mu k_2^\nu) ,
\end{align}
where we use $M_{a_0} = 980$ MeV, $M_{f_0} = 990$ MeV, and $\Gamma_{a_0,f_0} =
75$ MeV~\cite{Tanabashi:2018oca}.

The $\phi$-radiation invariant amplitudes are constructed as
\begin{align}
\label{eq:Amp:N}
\mathcal{M}_{\phi\,\mathrm{rad},s}^{\mu\nu} &= \frac{e g_{\phi NN}}{s-M_N^2}
\left( \gamma^\nu - i\frac{\kappa_{\phi NN}}{2M_N} \sigma^{\nu\alpha} k_{2\alpha}
\right) (\rlap{/}{q}_s + M_N)
\left( \gamma^\mu + i\frac{\kappa_N}{2M_N} \sigma^{\mu\beta} k_{1\beta} \right),
\cr
\mathcal{M}_{\phi\,\mathrm{rad},u}^{\mu\nu} &= \frac{e g_{\phi NN}}{u-M_N^2}
\left( \gamma^\mu + i\frac{\kappa_N}{2M_N} \sigma^{\mu\alpha} k_{1\alpha} \right)
(\rlap{/}{q}_u + M_N)
\left( \gamma^\nu - i\frac{\kappa_{\phi NN}}{2M_N} \sigma^{\nu\beta} k_{2\beta}
\right) ,
\end{align}
for the $s$ and $u$ channels, respectively.
$q_{t,s,u}$ are the four momenta of the exchanged particles, i.e.,
$q_t=k_2-k_1$, $q_s=k_1+p_1$, and $q_u=p_2-k_1$.

The invariant amplitudes for exchanges of spin-1/2 and -5/2 resonances are
computed as
\begin{align}
\label{eq:Amp:Ns}
\mathcal{M}_{N^*, 1/2^\pm}^{\mu\nu} &=
\frac{-ie}{s-M_{N^*}^2+i\Gamma_{N^*} M_{N^*}} \frac{h_{1}}{(2M_N)^2} 
\left[ g_1 \frac{M_\phi^2}{M_{N^*} \mp M_N} \Gamma^{\nu(\mp)} \mp
      ig_2 \Gamma^{(\mp)} \sigma^{\nu\beta} k_{2\beta} \right]
(\rlap{/}{q_s}+M_{N^*}) \Gamma^{(\mp)} \sigma^{\mu\alpha} k_{1\alpha},            \cr
\mathcal{M}_{N^*, 5/2^\pm}^{\mu\nu}
&= e \left[ \frac{g_1 }{(2M_N)^2}\Gamma_\rho^{(\mp)} 
      +\frac{g_2 }{(2M_N)^3} p_{2\rho} \Gamma^{(\mp)}  
      -\frac{g_3 }{(2M_N)^3} k_{2\rho} \Gamma^{(\mp)} \right]
k_2^{\beta_2} ( k_2^{\beta_1} g^{\nu\rho} - k_2^\rho g^{\nu \beta_1})
\Delta_{\beta_1 \beta_2 ; \alpha_1 \alpha_2}(q_s)                                  \cr
&\times
\left[ \frac{h_1}{(2M_N)^2} \Gamma_\delta^{(\mp)} \,\mp\,
\frac{h_2}{(2M_N)^3} \Gamma^{(\mp)} p_{1 \delta} \right]
k_2^{\alpha_2} (k_1^{\alpha_1} g^{\mu \delta} - k_1^\delta g^{\alpha_1 \mu}),   
\end{align}
for the $s$-channel diagram.
The propagator of a spin-5/2 baryon field is represented as~\cite{Chang:1967zzc}
\begin{align}
& \Delta_{\beta_2 \beta_1 ; \alpha_2 \alpha_1} (q) =
\frac{\rlap{/}{q}+M_{N^*}}{s-M_{N^*}^2+i\Gamma_{N^*} M_{N^*}}                  \\
& \times \left[\frac{1}{2}( \bar g_{{\beta_1}{\alpha_1}} \bar g_{{\beta_2}{\alpha_2}}
+\bar g_{{\beta_1}{\alpha_2}} \bar g_{{\beta_2}{\alpha_1}})
-\frac{1}{5} \bar g_{{\beta_1}{\beta_2}} \bar g_{{\alpha_1}{\alpha_2}}
-\frac{1}{10}( \bar \gamma_{\beta_1} \bar \gamma_{\alpha_1} \bar
g_{{\beta_2}{\alpha_2}}
+\bar \gamma_{\beta_1} \bar \gamma_{\alpha_2} \bar g_{{\beta_2}{\alpha_1}}
+\bar \gamma_{\beta_2} \bar \gamma_{\alpha_1} \bar g_{{\beta_1}{\alpha_2}} 
+\bar \gamma_{\beta_2} \bar \gamma_{\alpha_2} \bar g_{{\beta_1}{\alpha_1}} )
\right],                                                            \nonumber
\end{align}
with
\begin{equation}
\bar{g}_{\alpha\beta} = g_{\alpha\beta} - \frac{q_\alpha q_\beta}{M_{N^*}^2},
\,\,\,
\bar{\gamma}_\alpha = \gamma_\alpha - \frac{q_\alpha}{M_{N^*}^2}\rlap{/}{q}.
\end{equation}

\section*{Appendix B: Spin-Density Matrix Elements}
\label{AppenB}
The spin-density matrix elements (SDMEs) can be expressed in terms of the
helicity amplitudes~\cite{Schilling:1969um}:
\begin{align}
\label{eq:def:SDME}
\rho_{\lambda\lambda'}^0 &= \frac{1}{N} \sum_{\lambda_\gamma, \lambda_i, \lambda_f}
{\mathcal M}_{\lambda_f \lambda ; \lambda_i \lambda_\gamma}
{\mathcal M}_{\lambda_f \lambda'; \lambda_i \lambda_\gamma}^* ,                  \cr
\rho_{\lambda\lambda'}^1 &= \frac{1}{N} \sum_{\lambda_\gamma, \lambda_i, \lambda_f}
{\mathcal M}_{\lambda_f \lambda ; \lambda_i -\lambda_\gamma}
{\mathcal M}_{\lambda_f \lambda'; \lambda_i \lambda_\gamma}^* ,                  \cr
\rho_{\lambda\lambda'}^2 &= \frac{i}{N} \sum_{\lambda_\gamma, \lambda_i, \lambda_f}
\lambda_\gamma {\mathcal M}_{\lambda_f \lambda ; \lambda_i -\lambda_\gamma}
{\mathcal M}_{\lambda_f \lambda'; \lambda_i \lambda_\gamma}^* ,                  \cr
\rho_{\lambda\lambda'}^3 &= \frac{1}{N} \sum_{\lambda_\gamma, \lambda_i, \lambda_f}
\lambda_\gamma {\mathcal M}_{\lambda_f \lambda ; \lambda_i \lambda_\gamma}
{\mathcal M}_{\lambda_f \lambda'; \lambda_i \lambda_\gamma}^* ,
\end{align}
where the normalization factor $N$ is defined as
\begin{align}
\label{eq:def:NorFac}
N = \sum |{\mathcal M}_{\lambda_f \lambda ; \lambda_i \lambda_\gamma}|^2 .
\end{align}
The helicity states for the incoming photon and nucleon and the outgoing
nucleon are denoted by $\lambda_\gamma$, $\lambda_i$, and $\lambda_f$,
respectively, whereas $\lambda$ and $\lambda'$ stand for those for the
outgoing $\phi$ meson. By the symmetry property, the helicity amplitudes have
\begin{align}
\label{eq:TSymmeRela}
{\mathcal M}_{-\lambda_f -\lambda ; -\lambda_i -\lambda_\gamma} =
(-1)^{(\lambda-\lambda_f)-(\lambda_\gamma-\lambda_i)}
{\mathcal M}_{\lambda_f \lambda ; \lambda_i \lambda_\gamma} .
\end{align}
We have the following relations:
\begin{align}
\label{eq:RhoSymmeRela}
\rho_{\lambda\lambda'}^\alpha &= (-1)^{\lambda-\lambda'} \rho_{-\lambda-\lambda'}^\alpha
\,\,\,\,\, \mathrm{for} \,\,\, \alpha=0,1,                                \cr
\rho_{\lambda\lambda'}^\alpha &= -(-1)^{\lambda-\lambda'} \rho_{-\lambda-\lambda'}^\alpha
\,\,\,\,\, \mathrm{for} \,\,\, \alpha=2,3.
\end{align}

There is an ambiguity in choosing the quantization axis when computing the
SDMEs because they are not Lorentz-invariant quantities.
Thus the spin-quantization direction for the $\phi$ meson must be
determined.
We choose the Adair (A) frame, the helicity (H) frame, and the
Gottfried-Jackson (GJ) frame.
\begin{figure}[htp]
\centering
\includegraphics[width=9cm]{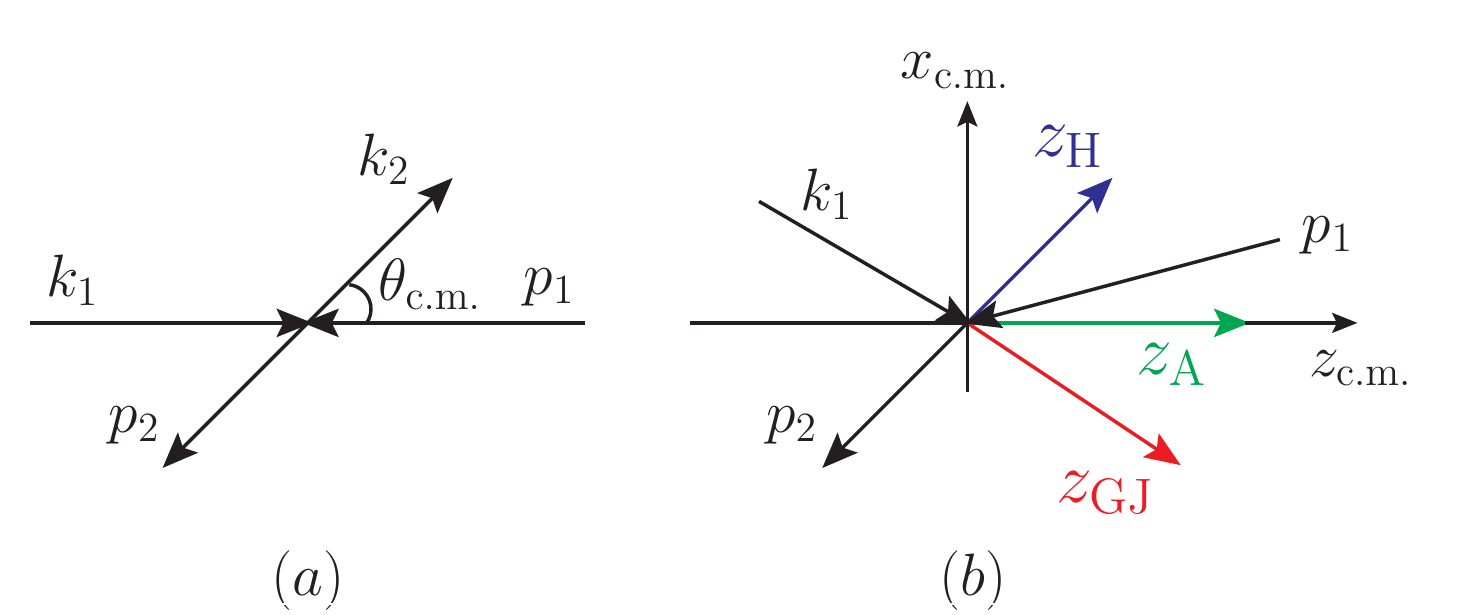}
\caption{Schematic diagrams for $\gamma p\to \phi p$ in (a)
the center-of-mass (c.m.) frame and (b) the $\phi$-meson rest frame.
A (green), H (blue), and GJ (red) stand for the Adair, helicity and
Gottfried-Jackson frames, respectively.}
\label{FIG:RefFrames}
\end{figure}
Figures~\ref{FIG:RefFrames}(a) and~\ref{FIG:RefFrames}(b) are schematic
diagrams in the c.m. frame and in the $\phi$-meson rest
frame, respectively.
In the Adair frame, the $z$ axis is parallel to the incoming photon
momentum in the c.m. frame.
The helicity and Gottfried-Jackson frames are when $z$ axis is
antiparallel to the momentum of the outgoing nucleon or is chosen to be
parallel to that of the incoming photon, respectively.
The former and latter ones are in favor of the $s$-channel and $t$-channel
helicity conservations, respectively.
When the SDMEs are given in one frame, it is straightforward to derive
them in other frames by a Wigner rotation.
The rotation angles are expressed as~\cite{Schilling:1969um,Dey:2014tfa}
\begin{align}
\label{eq: WignerAngle}
\alpha_{\mathrm{A}\to\mathrm{H}} &= \theta_{\mathrm{c.m.}} ,                     \cr
\alpha_{\mathrm{H}\to\mathrm{GJ}} &= -\cos^{-1}
\left( \frac{v-\cos\theta_{\mathrm{c.m.}}}{v\cos\theta_{\mathrm{c.m.}}-1}
\right),                                                                \cr
\alpha_{\mathrm{A}\to\mathrm{GJ}} &=
\alpha_{\mathrm{A}\to\mathrm{H}} + \alpha_{\mathrm{H}\to\mathrm{GJ}} ,
\end{align}
where $v$ is the velocity of the $K$ meson in the $\phi$-meson rest frame
(for the $\phi \to K \bar K$ decay).


\end{document}